\documentclass[floats,floatfix,showpacs,amssymb,prd,twocolumn,superscriptaddress,nofootinbib,reprint]{revtex4-2}

\usepackage{amssymb,amsmath,verbatim,mathtools,needspace,enumitem,etoolbox,graphicx,physics,microtype,afterpage,xspace,tabularx,lmodern,multirow,bm}
\usepackage{dcolumn}
\usepackage{multirow}
\usepackage{float}
\usepackage{booktabs}
\usepackage{gensymb}
\usepackage{lipsum}
\usepackage[dvipsnames, usenames]{xcolor}
\definecolor{linkcolor}{rgb}{0.0,0.3,0.5}
\usepackage[unicode, colorlinks=true, linkcolor=linkcolor, citecolor=linkcolor, filecolor=linkcolor, urlcolor=linkcolor, linktocpage, breaklinks]{hyperref}
\usepackage[all]{hypcap}
\usepackage[T1]{fontenc}
\usepackage[utf8]{inputenc}
\usepackage[usenames,dvipsnames]{xcolor}
\usepackage[normalem]{ulem}
\definecolor{TurkishBlue}{HTML}{144893}
\hypersetup{colorlinks=true,citecolor=TurkishBlue,linkcolor=TurkishBlue,urlcolor=TurkishBlue}

\definecolor{romared}{RGB}{142,0,28}

\newcommand{\be}{\begin{equation}}
\newcommand{\ee}{\end{equation}}

\def\be{\begin{equation}}
\def\ee{\end{equation}}
\newcommand{\beq}{\begin{eqnarray}}
\newcommand{\eeq}{\end{eqnarray}}

\usepackage{aas_macros}
\usepackage{makecell}
\usepackage{soul}
\usepackage[nolist,nohyperlinks]{acronym}

\usepackage{bm}
\renewcommand{\vec}{\bm}

\usepackage [english]{babel}
\usepackage [autostyle, english = american]{csquotes}
\MakeOuterQuote{"}

\newcommand{\approptoinn}[2]{\mathrel{\vcenter{
  \offinterlineskip\halign{\hfil$##$\cr
    #1\propto\cr\noalign{\kern2pt}#1\sim\cr\noalign{\kern-2pt}}}}}

\definecolor{darkgreen}{rgb}{0.33, 0.42, 0.18}

\newcommand{\lensMass}{M_{\rm Lz}}

\newcommand{\Msun}{M_{\odot}}
\newcommand{\orderOf}{\mathcal{O}}

\begin{document}

\title{Probing wave-optics effects and low-mass dark matter halos with\\lensing of gravitational waves from massive black holes}

\date{\today} 

\author{Mesut \c{C}al{\i}\c{s}kan}
\email{caliskan@jhu.edu}
\affiliation{William H.~Miller III Department of Physics and Astronomy, Johns Hopkins University, Baltimore, Maryland 21218, USA}

\author{Neha Anil Kumar}
\email{nanilku1@jhu.edu}
\affiliation{William H.~Miller III Department of Physics and Astronomy, Johns Hopkins University, Baltimore, Maryland 21218, USA}

\author{Lingyuan Ji}
\affiliation{Department of Physics, University of California, Berkeley, 366 Physics North MC 7300, Berkeley, California 94720, USA}

\author{Jose M.~Ezquiaga}
\affiliation{Niels Bohr International Academy, Niels Bohr Institute, Blegdamsvej 17, DK-2100 Copenhagen, Denmark}

\author{Roberto Cotesta}
\affiliation{William H.~Miller III Department of Physics and Astronomy, Johns Hopkins University, Baltimore, Maryland 21218, USA}

\author{Emanuele Berti}
\affiliation{William H.~Miller III Department of Physics and Astronomy, Johns Hopkins University, Baltimore, Maryland 21218, USA}

\author{Marc Kamionkowski}
\affiliation{William H.~Miller III Department of Physics and Astronomy, Johns Hopkins University, Baltimore, Maryland 21218, USA}

\begin{abstract}
\noindent 
The Laser Interferometer Space Antenna (LISA) will detect gravitational waves (GWs) emitted by massive black hole binaries (MBHBs) in the low-frequency ($\sim$mHz) band.
Low-mass lenses, such as low-mass dark matter halos or subhalos, have sizes comparable to the wavelength of these GWs.
Encounters with these lenses produce wave-optics (WO) effects that alter waveform phase and amplitude. 
Thus, a single event with observable WO effects can be used to probe the lens properties.
In this paper, we first compute the probability of observing WO effects in a model-agnostic way. 
We perform information-matrix analyses over $\mathcal{O}(1000)$ MBHBs with total mass, mass ratio, and redshift spanning the ranges relevant to LISA. 
We then calculate lensing rates using three semi-analytical models of MBHB populations. 
In both cases, we use a waveform model that includes merger, ringdown, and higher-order modes.
We use two lens population models: the theory-based Press-Schechter halo mass function and an observation-based model derived from Sloan Digital Sky Survey.
We find that the probability of detecting WO effects can be as large as $\sim 3\%$, $\sim1.5\%$, and $\sim 1 \%$ at $1\sigma$, $3\sigma$, and $5\sigma$ confidence levels, respectively.
The most optimistic MBHB population model yields $\sim 8$, $\sim 4$, and $\sim 3$ events with detectable WO effects at the same confidence levels, while the rates drop to $\sim 0.01$ in the more pessimistic scenarios.
The most likely lens masses probed by LISA are in the range $(10^3, 10^8)\, M_{\odot}$, and the most probable redshifts are in the range $(0.3, 1.7)$.
Therefore, LISA observations of WO effects can probe low-mass DM halos, complementing strong lensing and other observations.
\end{abstract}

\maketitle

\section{Introduction} \label{sec:Introduction}

The large-scale structure of the Universe encodes invaluable information about cosmic history and its fundamental constituents. 
Of particular importance is dark matter (DM), whose properties are key to understanding the formation of the cosmic web~\cite{Springel:2005nw, Planck:2018vyg}. 
By mapping the distribution of galaxies across the sky one can learn about the DM halos that host them and their assembly history.
However, it is difficult to probe their substructure, as most of the subhalos are barely luminous, and the optical depths associated with them are low due to their minuscule masses~\cite{Bullock:2000wn, Moore:1999nt}.
A theoretical understanding of the low-mass DM halo population can still be achieved with analytical calculations, such as the Press-Schechter (PS) formalism~\cite{Press:1973iz}, or numerical simulations~\cite{Sheth:1999su, Tinker:2008ff}, allowing predictions on the abundances of such objects for different DM models. 
The small-scale structure is suppressed in models in which the DM is warmer than the standard cosmological model~\cite{Bode:2000gq, Viel:2013fqw}.

Because everything in the Universe is subject to gravity, gravitational lensing (hereafter lensing) offers a unique opportunity to probe the dark side of cosmological structures. For example, light emitted by distant galaxies will encounter inhomogeneities that act as lenses, deflecting their trajectories, and altering their shapes~\cite{Tyson:1998vp, Bartelmann:1999yn, Refregier:2003ct}.
When the source and lens are well aligned, the effect of lensing becomes stronger and produces multiple images of the same source~\cite{Schneider:1992, Bartelmann:2010fz, Narayan:1996ba}. If the images are highly magnified, it may be possible to zoom into strongly lensed galaxies and thus explore the small perturbations induced by the substructure.
However, current techniques and observational capabilities are limited to DM halos more massive than $\sim10^9\, M_\odot$~\cite{Hezaveh:2016, Sengul:2021lxe, Sengul:2023olf}.

When distant galaxies merge, their central black holes can get close enough so that gravitational radiation drives them to coalescence. The gravitational waves (GWs) emitted by these merging massive black hole binaries (MBHBs) will be one of the primary targets for the future Laser Interferometer Space Antenna (LISA)~\cite{2017arXiv170200786A}. 
Because of their cosmological distances---most mergers are expected within $2\lesssim z \lesssim8$---there is a chance that the GWs themselves can be lensed~\cite{Sereno:2010dr, Takahashi:2003ix, Hannuksela:2020xor,Nakamura:1997sw, Wang:1996as,Mukherjee:2019wcg, Caliskan:2022wbh}.
When the lenses are large (for example galaxies, groups of galaxies, or clusters), the phenomenology of LISA GWs is similar to that of electromagnetic radiation: the so-called geometric-optics approximation is valid, and these GWs can be weakly or strongly lensed~\cite{Mizuno:2022xxp, Sereno:2011ty, Takahashi:2005sxa, Wu:2022vrq, LIGOScientific:2023bwz}. 
In the latter case, echoes of the original coalescence are produced.

In contrast, when lenses are smaller so that the wavelength of GWs becomes comparable to the characteristic size of the lens, a new phenomenology emerges in the form of wave-optics (WO) effects~\cite{Ohanian:1974ys,Deguchi:1986zz,Takahashi:2003ix}.
In the WO regime, the GW signal becomes distorted due to \textit{frequency-dependent} modulations in the waveform phase and amplitude.
Lensing can be identified by tracking these changes with respect to the unlensed waveform templates, which means that the lens mass and lens density profile can be constrained using single waveforms.\footnote{The observation of SL is affected by the possible occurrence of false alarms, caused by the intricate task of correctly identifying two or more different waveforms as being lensed ``copies'' of the same source~\cite{Caliskan:2022wbh}.
The use of WO effects on a single waveform to probe lens properties has the notable advantage of mitigating this issue, which can also be alleviated by the observation of a Type-II image with detectable higher-order modes~\cite{Ezquiaga:2020gdt}.}

This implies that the cross section of lensing detectability is not limited to the region of the sky where the source would be close enough to the lens to form multiple images, and the probability of detecting WO can be considerably higher than naively extrapolated from strong lensing (SL)~\cite{Takahashi:2003ix, Caliskan:2022hbu, Xu:2021bfn, Ezquiaga:2020gdt, Sereno:2010dr}. 
Consequently, LISA can probe the nature of DM halos and subhalos in a regime that is difficult to probe with current techniques.

Several recent studies have focused on WO effects~\cite{Ezquiaga:2020spg,Cremonese:2021ahz,Garoffolo:2022usx, Gao:2021sxw, Tambalo:2022plm, Fairbairn:2022xln, Tambalo:2022wlm, Savastano:2023spl, Jung:2022tzn, Guo:2022dre}.
Through the first information-matrix analysis with waveform models incorporating the merger, ringdown, higher harmonics, and aligned spins, we demonstrated that the higher-order modes and the merger/ringdown part of the waveform significantly enhance LISA's lens parameter measurement capability~\cite{Caliskan:2022hbu} relative to estimates in the earlier literature, such as Ref.~\cite{Takahashi:2003ix}.

Recent work calculated the probability of detection of WO effects for LISA, finding a peak at approximately $5 \times 10^{-5}$~\cite{Fairbairn:2022xln}.
The study employed the lowest-order approximations for both the amplitude and phase of the binary, thus characterizing the waveform with only four parameters and considered inspiral-only waveforms truncated at the innermost stable circular orbit.
Their finding of a low lensing probability can be attributed to the fact that inspiral-only waveforms underestimate LISA's lens parameter detection capability~\cite{Caliskan:2022hbu}.
Another recent study~\cite{Savastano:2023spl} also explored the probability of WO effect detection using inspiral-merger-ringdown phenomenological waveforms for the leading quadrupolar radiation and found a very different result, with a peak at approximately $20\%$. 
However, the authors of Ref.~\cite{Savastano:2023spl} employed the so-called Lindblom criterion~\cite{Lindblom:2008cm}, an arguably optimistic metric that can neglect parameter degeneracies. The Lindblom criterion is known to overestimate lens parameter measurement capabilities, and thus detection probabilities~\cite{Caliskan:2022hbu}. 

While such studies offer valuable insights into the WO-effect detection probability and the potential to probe different DM scenarios, 
their methodology could be improved by more comprehensive waveform models,
the full LISA detector response, and
detection criteria that go beyond waveform mismatches.
Furthermore, it is important to carry out an extensive exploration of the influence of mass, mass ratio, redshift, and other source parameters on the WO detection probability.

In this study, we calculate the likelihood of observing WO effects through two distinct approaches:
\begin{enumerate}
    \item A model-agnostic exploration of approximately $1000$ MBHBs spanning the total mass, mass ratio, and redshift ranges relevant to LISA;
    \item A study of three representative semi-analytical astrophysical models of MBHB populations.
\end{enumerate}
The first approach highlights how the WO optical depth and detection probability depend on the intrinsic MBHB source parameters (and in particular on mass, mass ratio, and redshift). The second approach provides specific estimates for the lensing rates and shows how the global characteristics of the MBHB population influence detection prospects.

Our information-matrix formalism uses inspiral-merger-ringdown waveform models including aligned spins and higher-order modes. 
It accounts for potential degeneracies between the source and lens parameters and considers the full detector response of LISA. We introduce a new and much faster parameter estimation method based on lookup tables, which enables us to investigate the lensing likelihood for $\orderOf(2000)$ MBHBs in total.
We take into account the effect of different detection thresholds, using both theoretically and observationally motivated lens populations.
Two of the most interesting outcomes of our work are predictions of the halo-mass and redshift ranges that can be probed by LISA using WO effects.

The paper starts with a brief discussion of lensing of GWs and parameter estimation in Sec.~\ref{sec:Parameter_Estimation}, with further details in Appendix~\ref{App:LUT}.
In Sec.~\ref{sec:Lens_Population}, we introduce our lens populations and the rationale behind their choice.
In Sec.~\ref{sec:Source_Population}, we detail the model-agnostic approach and the semi-analytical models used to generate astrophysical populations of MBHBs.
In Sec.~\ref{sec:Probability}, we detail the computation of the probability of observing WO effects.
After this, we extensively explore the probability of WO effects considering different detection thresholds.
In Sec.~\ref{sec:results_model_agnostic}, we present the results for the model-agnostic approach, providing an extensive overview of lensing probabilities for MBHBs detectable by LISA.
In Sec.~\ref{sec:results_pop_model}, we present the results (including lensing rates) for three representative MBHB source populations.
Then, in Secs.~\ref{sec:results_lens_mass} and~\ref{sec:results_lens_redshift}, we present the range of halo masses and halo redshifts that can be probed with the observation of WO effects.
Finally, in Sec.~\ref{sec:Conclusions}, we summarize our conclusions and highlight potential directions for future work.

Throughout the paper, we assume a $\Lambda$CDM cosmology with cosmological parameter values matching Planck 2018~\cite{Planck:2018vyg}, as implemented by \texttt{astropy}~\cite{2013astropy, 2018astropy, 2022astropy}. Unless specified otherwise, we work in geometric units $(G=c=1)$.

\section{Gravitational Lensing and Parameter Estimation} \label{sec:Parameter_Estimation}

We are interested in the merger of MBHBs with primary and secondary (source-frame) masses $m_1$ and $m_2$ at source redshift $z_{\rm S}$. For each binary, it is convenient to introduce the source-frame total mass $M_{\rm T}=m_1+m_2$, the mass ratio $q=m_1/m_2 \geq 1$, and the detector-frame (redshifted) total mass $M_{\rm Tz} = (1+z_{\rm S})M_{\rm T}$.
The lensed gravitational waveform $\tilde h^{\rm L}(f; \bm \theta^{\rm L}, \bm \theta^{\rm S}) \equiv \tilde h^{\rm L}_{+} - i\tilde h^{\rm L}_{\times}$ in the frequency domain is given by the product of the ``diffraction integral'' $F(f;\bm \theta^{\rm L})$, whose detailed expression is given in Eq.~\eqref{eq:diffraction_integral} below, and the unlensed waveform:
\begin{equation}
    \tilde h^{\rm L}(f; \bm \theta^{\rm L}, \bm \theta^{\rm S}) = F(f;\bm \theta^{\rm L})\tilde h(f; \bm \theta^{\rm S})\, .
\end{equation}
Here, $f$ is the GW frequency, and the 11-dimensional vector $\bm{\theta}^{\rm S} \equiv \{M_{\rm Tz},\ q,\ d_{l},\ t_{\rm c},\ \iota,\ \phi_{\rm c},\ \lambda,\ \beta,\ \psi,\ \chi_{\rm m},\ \chi_{\rm p} \}$ includes the following source parameters: the detector-frame total mass $M_{\rm Tz}$, mass ratio $q$, luminosity distance to the source $d_{l}$, coalescence time $t_{\rm c}$, inclination angle $\iota$, coalescence phase $\phi_{\rm c}$, right ascension $\lambda$, declination $\beta$, polarization angle $\psi$, and two parameters---the ``effective spin'' $\chi_{\rm p}=(m_1\chi_1+m_2\chi_2)/(m_1+m_2)$ and the asymmetric spin combination $\chi_{\rm m}=(m_1\chi_1-m_2\chi_2)/(m_1+m_2)$---for the spins of the binary components, which we assume to be aligned with the orbital angular momentum.
The vector $\bm{\theta}^{\rm L}$ includes, in contrast, the lens parameters, which depend on the chosen model for the lens.

The diffraction integral for a given $f$ reads
\begin{equation}\label{eq:diffraction_integral}
    F(f, \vec y) \equiv \frac{D_\mathrm{S}(1+z_\mathrm{L})\xi^2_0}{D_\mathrm{L} D_\mathrm{LS}} \frac{f}{i} \int \mathrm{d}^2 x\, \exp[2\pi i f t_d(\vec x, \vec y)],
\end{equation}
where the lensing time delay is defined as
\begin{equation}
    t_\mathrm{d}(\vec x, \vec y) \equiv \frac{D_\mathrm{S} \xi^2_0}{D_\mathrm{L} D_\mathrm{LS}} (1 + z_\mathrm{L}) \left[\frac12 |\vec x - \vec y|^2 - \psi(\vec x) + \phi(\vec y)\right].
\end{equation}
Here, $D_\mathrm{L}$, $D_\mathrm{S}$, and $D_\mathrm{LS}$ are the angular-diameter distances to the lens, to the source, and from the lens to the source, respectively. 
The dimensionless 2-vectors $\vec x \equiv \vec \xi / \xi_0$ and $\vec y \equiv \bm{\eta} D_\mathrm{L} / (D_\mathrm{S}\xi_0)$ are the normalized versions of the image-plane coordinates $\vec \xi$ and the source-plane coordinates $\vec \eta$, while $\xi_0$ is the Einstein radius of the lens. 
The properties of the lens are encoded in the lensing potential $\psi(\vec x)$, and $\phi(\vec y)$ is an $\vec x$-independent constant chosen so that $\min_{\vec x} t_\mathrm{d}(\vec x, \vec y) = 0$. For axially symmetric lenses, all 2-vectors ($\vec x, \vec y, \dots$) are reduced to scalars ($x, y, \dots$). Sometimes it is convenient to use the dimensionless frequency
\begin{equation}
    w \equiv \frac{D_\mathrm{S} \xi^2_0 (1+z_\mathrm{L})}{D_\mathrm{L} D_\mathrm{LS}} 2\pi f
\end{equation}
instead of the physical frequency $f$.

In this paper, we focus on the singular isothermal sphere (SIS) lens, thus following the treatment detailed in Ref.~\cite{Caliskan:2022hbu} (see, specifically, Sec.~II~B and Appendix A therein).
This class of lens models is straightforwardly parametrized using the lens parameters $\bm \theta^{\rm L} \equiv \{ M_{\rm Lz},\, y \}$, where $M_{\rm Lz}=(1+z_\mathrm{L})M_{\rm L}$ is the redshifted lens mass, and $y$ is the so-called impact parameter.
For an SIS lens, the lens mass is related to the Einstein radius $\xi_0$ and dimensionless frequency $w$ as follows:
\begin{align}\label{eqn:Einstein_rad_and_w}
    \xi_0 = \left (\frac{4 M_{\rm L} D_{\rm L} D_{\rm LS}}{D_{\rm S}}\right)^{1/2} && \textrm{and} && w = 8\pi M_{\rm Lz} f\,.
\end{align}

We follow the information-matrix analysis (or linear signal approximation) described in Ref.~\cite{Caliskan:2022hbu} to determine the uncertainties in estimating the parameters of the lens and MBHB system.
Our 13-dimensional matrices include the 11 source parameters as well as the lens parameters and account for any possible degeneracies between them.

An accurate calculation of the probability of observing WO requires scanning the lens parameter space $(\lensMass,\, y)$ and considering all potential lensing configurations that can produce measurable effects. 
The most time-consuming part of the information-matrix analysis is solving the lensing diffraction integral $F(w, y)$ and its derivatives at each $f$. 
We will investigate the probability of observing WO for $\orderOf(2000)$ MBHBs. This requires $\orderOf(10^7)$ information matrix calculations and, therefore, it is essential to speed up the evaluation of $F(w, y)$ and its derivatives.
To achieve this aim, we create a lookup table that allows us to rapidly compute $F(w, y)$ and its derivatives with respect to 
$y \in \left[0.01, 300\right]$ and $f \in \left[10^{-5}, 0.5 \right] \textrm{Hz}$. 
Our lookup table method speeds up the analysis by a factor $\mathcal{O}(10^3\, \textrm{--}\, 10^4)$ while maintaining the same level of accuracy reported in~\cite{Caliskan:2022hbu}.
A more detailed discussion of the lookup table can be found in Appendix~\ref{App:LUT}.

\section{Lens Populations} \label{sec:Lens_Population}

The population of lenses to be accounted for in a given source-observer configuration is defined by the number of massive, virialized objects in the line of sight of the source. 
Various approaches have been used to characterize the mass dependence and redshift evolution of the lens population, across theory, simulation, and observation~\cite{Press:1973iz, Sheth:1999su, Tinker:2008ff, Mitchell:2004gw}. 
However, there still remain vast uncertainties in the expected number densities of these objects. These uncertainties are particularly apparent in the high-redshift, low-mass regime, where lenses are not luminous enough to be detected directly and not massive enough to strongly lens background sources.
In this work, through the use of multiple lens populations, we show that measurements of the lensing of GWs in the wave-optics regime can be used to effectively constrain the population of lenses, particularly in the low-mass and high-redshift regimes.

One population of lenses we consider is derived purely from the theoretical approach of modeling the collapse of virialized objects. This model was first explored in Ref.~\cite{Press:1973iz} and is called the PS halo mass function. Although modifications have been made to this initial analysis based on large-scale cosmological simulations (see Refs.~\cite{Sheth:1999su, Tinker:2008ff}), we choose to adhere to the simplest, theoretically motivated model due to the persistence of uncertainties across all the proposed models, as well as the similarity in expected rates across PS and newer simulation-based modifications. It is important to note, however, that this model maps \textit{all} virialized objects using a fixed over-density threshold, treating each collapse as an isolated event.
As a result, it predicts a very large number of halos in the low-mass regime that has been very difficult to observationally constrain~\cite{Press:1973iz, Ishiyama:2014uoa, Jenkins:2000bv, Lacey:1994su}. 
Therefore, we view the rates calculated from this lens population as a potentially optimistic upper bound on the expected number of lensed events. 

The second population of lenses explored in our analysis is motivated by observations of the Sloan Digital Sky Survey (SDSS), called the \textit{measured velocity function} (MVF). 
This model, first explored in Ref.~\cite{Mitchell:2004gw}, characterizes the number density of virialized objects as a function of their observed velocity dispersion. 
This model was initially developed within the context of strong-lensing measurement statistics. As a result, the resulting fit to observational data is derived mainly from the population of massive, early-type galaxies.
This indicates that this model can be seen as a robust description of the number density of higher-mass galaxies with virial mass $\sim 10^{13}- 10^{14}M_{\odot}$. 
However, when considering the extension of this model to the lower-mass regime, one must note that lower-mass halos are less luminous and less likely to act as strong lenses for background sources. Therefore, this observation-based model should be considered as a conservative, but robust lower bound on the expected number of lensed events. 

This section aims to introduce the notation and detail the two different populations of lenses used in our analysis. 
Furthermore, we assume that all lenses follow the SIS profile~\cite{Schneider:1992, Narayan:1996ba}, and go on to derive the analytic relations between lens characteristics, such as the lens mass and Einstein radius, and general halo characteristics, such as virial mass and velocity dispersion. 
Recent work has explored the possibility of probing different lens profiles for halos through their distinct WO effects~\cite{Cremonese:2021ahz,Fairbairn:2022xln,Savastano:2023spl}.

\begin{figure}[t]
\includegraphics[width=\columnwidth]{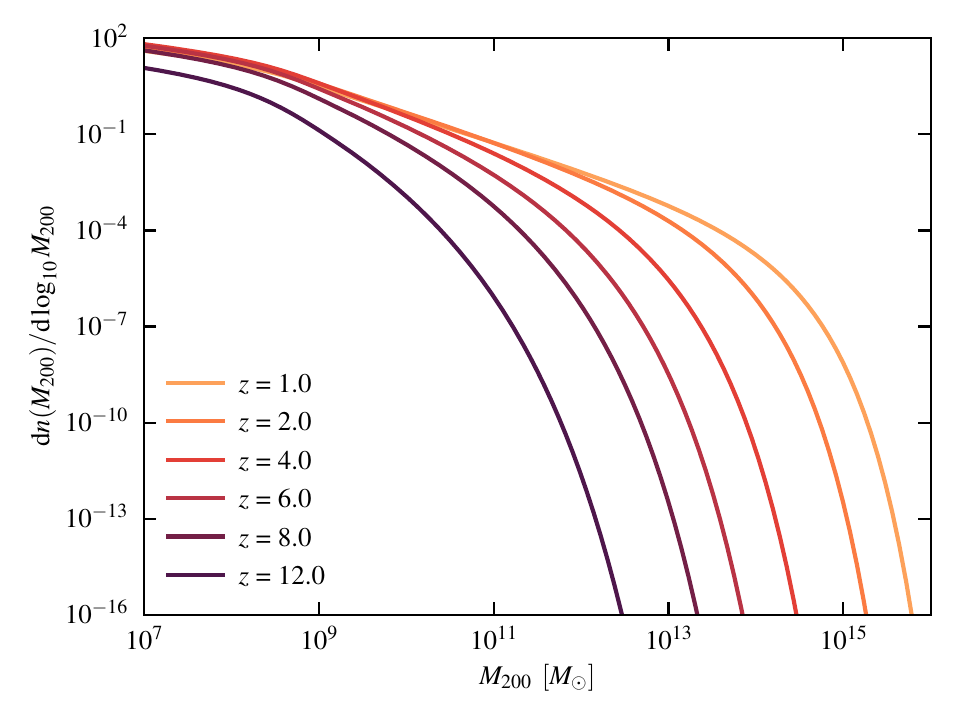} 
\caption{Differential halo mass function per $\log_{10}$ mass bin as a function of halo (virial) mass $M_{200}$ for various halo (lens) redshifts $z$. Here, we assume a PS model for the collapse fraction~\cite{Press:1973iz}.
The plot demonstrates how, as the halo redshift decreases, more of the lower mass halos have accreted to higher masses.
}
    \label{fig:hmf_redshifts}
\end{figure}

\begin{figure}[t]
    \includegraphics[width=\columnwidth]{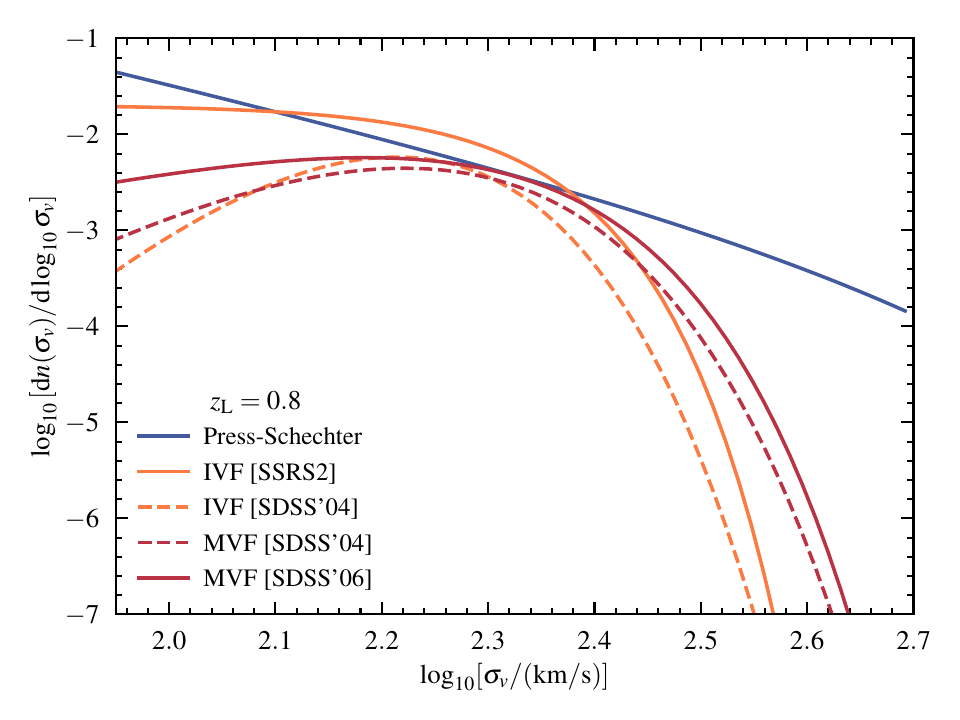} 
    \caption{Velocity function $\mathrm{d}n(\sigma_v)/\mathrm{d}\log_{10}{\sigma_v}$ as a function of velocity dispersion $\sigma_v$ for 5 different lens populations.
    The blue line shows the velocity function calculated using the Press–Schechter (PS) halo mass function~\cite{Press:1973iz}.
    The solid and dashed orange lines are the inferred velocity functions from \texttt{SSRS2}~\cite{2019PASA...36...33O} and \texttt{SDSS'04}~\cite{SDSS:2004wzw, SDSS:2004ulj}, respectively.
    The dashed and solid red lines are the measured velocity function based on \texttt{SDSS'04}~\cite{Mitchell:2004gw} and \texttt{SDSS'06}~\cite{Choi:2006qg}, respectively.
    The values are shown for lens populations at a lens redshift $z_{\rm L}=0.8$, which is approximately the horizon distance of the SDSS data set.
    In our analysis, we use the PS and MVF [\texttt{SDSS'06}] models to characterize the lens populations.
}
    \label{fig:vel_func_comparison}
\end{figure}

\subsection{The Press-Schechter Halo Mass Function} \label{sec:PS_halo_mass_function}
The halo mass function $n(M, z)$ represents the differential number density of halos at a given virial mass $M$ and redshift $z$, i.e., $n(M,z)\ \mathrm{d}M$ defines the number of halos, per unit comoving volume, with masses in the range $M \rightarrow M + \mathrm{d}M$ at redshift $z$. This function has a simple analytic form, given by
\begin{equation}
    n(M, z)\ \mathrm{d} M = \frac{\rho_m}{M^2}f(\sigma,z)\frac{\mathrm{d}\ln \sigma(M, z)}{\mathrm{d}\ln M}\ \mathrm{d}M,
\end{equation}
where $M$ represents the virial mass of a halo, $\rho_m$ is the present-day cosmological matter density, and $f(\sigma, z)$ is conventionally called the ``collapse-fraction.'' The above equation also depends on the rms variance of mass within a sphere of radius $R$, defined as
\begin{equation}
    \sigma^2(M, z) = \frac{1}{2\pi^2}\int_{0}^{\infty} \mathrm{d}k \ k^2\  P_{mm}(k,z)W^2(kR),
\end{equation}
where $R = [3M/(4\pi\rho_m)]^{1/3}$, and $P_{mm}(k,z)$ is the linear matter power spectrum, with each mode weighted by the window function in Fourier space, given by:
\begin{equation}
    W(kR)  = \frac{3[\sin(kR) - kR\cos(kR)]}{(kR)^3}.
\end{equation}

The equations so far represent a model-independent description of the distribution of halos on large scales. The model dependency of this formalism arises with the choice of collapse fraction $f(\sigma, z)$. Despite the existence of more recent, simulation-based computations of this fraction (see e.g.~\cite{Jenkins:2000bv, Tinker:2008ff}), given the uncertainty in these distributions, we choose to use the analytic PS model introduced in Ref.~\cite{Press:1973iz}:
\begin{equation}
    f(\sigma, z) = \sqrt{\frac{2}{\pi}}\frac{\delta_c}{\sigma(M, z)}\exp\left(-\frac{\delta_c^2}{2\sigma(M,z)^2}\right),
\end{equation}
with $\delta_c = 1.686$. It is important to note that this formalism describes the halo distribution in terms of virial mass $M_{200}$, where the explicit subscript has been suppressed for ease of notation. This quantity is defined as the total mass enclosed within the virial radius $r_{200}$ of the halo in consideration. The subscript indicates that $r_{200}$ is the radius at which the density of the system is 200 times the critical density of the universe at the given redshift. Figure \ref{fig:hmf_redshifts} shows the PS halo mass function at various redshifts, as a function of halo virial mass. 

\subsection{The Measured Velocity Function} \label{sec:Measured_velocity_function}
The velocity function $n(\sigma_v, z)$ describes the differential number density of virialized objects with velocity dispersion $\sigma_v$ at redshift $z$. Given a model for the distribution of mass within the object, the velocity dispersion has a one-to-one relation with the virial mass of the halo.

The model we use in our analysis, called the MVF, was first explored in Ref.~\cite{Mitchell:2004gw} using data from SDSS in 2004 (\texttt{SDSS'04})~\cite{SDSS:2004wzw, SDSS:2004ulj}. Prior to this analysis, $n(\sigma_v)$ was obtained from galaxy surveys by combining the measured galaxy luminosity function with the empirical Faber-Jackson relation. Consistently with previous work, we call this the \textit{inferred velocity function} (IVF). However, this estimator for $n(\sigma_v)$ results in a biased distribution due to the scatter in the Faber-Jackson relation and uncertainties in the redshift evolution of the luminosity function~\cite{Mitchell:2004gw}. Therefore, the authors of Ref.~\cite{Mitchell:2004gw} formulated the MVF by directly fitting the measurements from SDSS to a modified Schechter function as follows:
\begin{eqnarray}
    n(\sigma_v)\ \mathrm{d}\sigma_v = \phi_* \left(\frac{\sigma_v}{\sigma_*}\right)^\alpha e^{-\left(\frac{\sigma_v}{\sigma_*}\right)^\beta}\frac{\beta}{\Gamma(\alpha/\beta)}\frac{\mathrm{d}\sigma_v}{\sigma_v},
\end{eqnarray}
where $\{\phi_*, \sigma_*, \alpha, \beta\}$ are the fitting parameters. 

Redshift evolution can be included, as discussed in Ref.~\cite{Mitchell:2004gw}, by varying the parameter $\phi_*$. However, many studies on strong-lensing statistics have assumed the velocity function to be constant in comoving units. This is because the catalogs used to construct these mass-distribution functions are dominated by massive, early-type galaxies that are expected to evolve only passively through their luminosity functions~\cite{Im:2000vi, Ofek:2003sp}. Given the fact that these models are based on local, observational data, we choose not to complicate the analysis with a redshift dependence extending to $z \sim 10$. Instead, we adhere to the simplifying assumption of a constant number density in velocity-dispersion space. It is important to note that
the number-density of lower-mass, early-type galaxies is expected to increase from $z = 0$ to $z = 1$, while the number density of the higher-mass halos decreases over the same interval~\cite{Mitchell:2004gw}.
This indicates that our current set-up of a non-evolving velocity function can still be considered a conservative lower bound on lensing rates.

In our analysis we use the updated set of fitting parameters measured in Ref.~\cite{Choi:2006qg} using \texttt{SDSS'06}~\cite{SDSS:2007aih}:
\begin{equation}
\begin{split}
    \{\phi_*, \sigma_*, & \alpha, \beta\} = \\
    & \{8.0 \times 10^{-3} h^3 {\rm{Mpc}}^{-3},  161\ {\rm km\ s}^{-1}, 2.32, 2.67\}\,.
\end{split}
\end{equation}
These fitting parameters are derived from measurements of galaxies with redshift $0.025 \lesssim z \lesssim 0.1$ and velocity dispersion $70\ {\rm km\ s}^{-1} \lesssim \sigma_v \lesssim 300\ {\rm km\ s}^{-1}$.

In Fig.~\ref{fig:vel_func_comparison}, we plot the two different MVFs measured from \texttt{SDSS'04} in Ref.~\cite{Mitchell:2004gw} and \texttt{SDSS'06} in Ref.~\cite{Choi:2006qg}. Furthermore, to compare these directly measured velocity functions to the inferred scenario derived from the measured luminosity function, we also display two different IVFs measured from \texttt{SDSS'04} and the Second Southern Sky Survey (\texttt{SSRS2}) in Ref.~\cite{Mitchell:2004gw}. When plotted against the PS halo mass function, it is clear that this distribution, calculated solely based on measurements of early-type galaxies, predicts a very small number of objects at lower masses. Note that the plot only depicts a limited range of velocity dispersions, corresponding to similar plots made in Refs.~\cite{Mitchell:2004gw, Choi:2006qg}. 
However, in our analysis, we extrapolate the model to include lenses of lower mass and lower velocity dispersion. 

\subsection{Tying the Mass Function to SIS Density Profile} \label{sec:mass_function_to_profile}

For this work, we assume that all lenses are characterized by the SIS mass-density profile~\cite{Schneider:1992, Narayan:1996ba}. 
In this subsection, we analytically connect the virial mass of a halo to its Einstein radius and lens mass for a particular source-lens-observer configuration under this assumption. 

We begin by introducing the 3-dimensional density profile:
\begin{equation}
    \rho_{\rm{SIS}}(r) = \frac{\sigma_v^2}{2\pi G r^2},
\end{equation}
where $r$ represents the 3D radius from the center of the halo.
By integrating this density profile along the 3D radius we find the total mass contained within $r$,
\begin{equation}
    M_{\rm SIS} (<r) = \frac{2\sigma_v^2 r}{G}, 
\end{equation}
which allows us to write the virial mass $M_{200}$ for an SIS halo as a function of $r_{200}$.
Similarly, we can integrate along the line of sight to find the projected surface matter density 
\begin{equation}
    \Sigma_{\rm{SIS}}(R) = \frac{\sigma_v^2}{2GR},
\end{equation}
where $R$ is the projected 2D radius from the halo center. Therefore, the mass enclosed within the projected radius $R$ is
\begin{equation}
    M_{\rm{SIS}} (<R) = \frac{\sigma_v^2}{2G} R. 
    \label{eq: 2D Mass in R}
\end{equation}
The lens mass and Einstein radius are not only dependent on the assumed mass-density profile of the lens, but also on the source-lens-observer configuration. Using the above equations, for a given cosmology and lensing geometry, one can define the critical density as~\cite{Robertson:2020mfh}
\begin{equation}
    \Sigma_{\rm crit} = \frac{c^2}{4\pi G}\frac{D_{\rm S}}{D_{\rm L} D_{\rm LS}}.
\end{equation}
The Einstein radius can then be defined in terms of this critical density as:
\begin{equation}
    \xi_0 = \frac{\sigma_v^2}{G\Sigma_{\rm crit}D_{\rm L}}.
    \label{eq: EinsteinRadius}
\end{equation}
Therefore, based on Eqs.~\eqref{eq: 2D Mass in R} and~\eqref{eq: EinsteinRadius}, we can write the lens mass for an SIS lens in terms of $\sigma_v$ as
\begin{equation}
    M_{\rm L} = \frac{4\pi^2 \sigma_v^4}{c^2 G} \frac{D_{\rm L} D_{\rm LS}}{D_{\rm S}}.
\end{equation}
Finally, the virial mass of an SIS halo is related to its lens mass for a given source-lens-observer configuration as follows:
\begin{equation}
    M_{\rm L} = \frac{\pi^2 G D_{\rm L} D_{\rm LS}}{c^2 D_{\rm S}}\Bigg( \frac{M_{200}}{r_{200}}\Bigg)^2\,.
\end{equation}\\
For reference, a plot representing the above relation between $M_{\rm L}$ and $M_{200}$ for different choices of $z_{\rm L}$ and $z_{\rm S}$ can be found in Appendix~\ref{App:lens_mass_vs_halo_mass}.

\section{Source Populations}\label{sec:Source_Population}

The likelihood of observing lensing is highly dependent on the source properties of MBHBs~\cite{Caliskan:2022hbu}.
This implies that apart from variations in the optical depth due to changes in $z_{\rm S}$, the probability of observing lensing will also be significantly influenced by the measurable range of $M_{\rm Lz}$ and $y$, which are in turn determined by the parameters of the MBHBs.
Furthermore, considering the differing distributions of their parameters (total mass, mass ratio, redshift, and so on) and their merger rates, a model for the source population is necessary to compute a lensing detection rate.
For a comprehensive understanding of lensing probabilities and rates, it is necessary to investigate the effects of different populations and to examine the probabilities over wide ranges of source parameters.

In this section, we explore both a model-agnostic case (Sec.~\ref{sec:source_model_indep}), in which we scan the parameter space relevant for LISA observations using approximately 1000 MBHBs, and a scenario based on three semi-analytical population models often used in the literature (Sec.~\ref{sec:source_pop_model}).

\begin{figure*}[t]
    \centering
    \includegraphics[width=.99\textwidth]{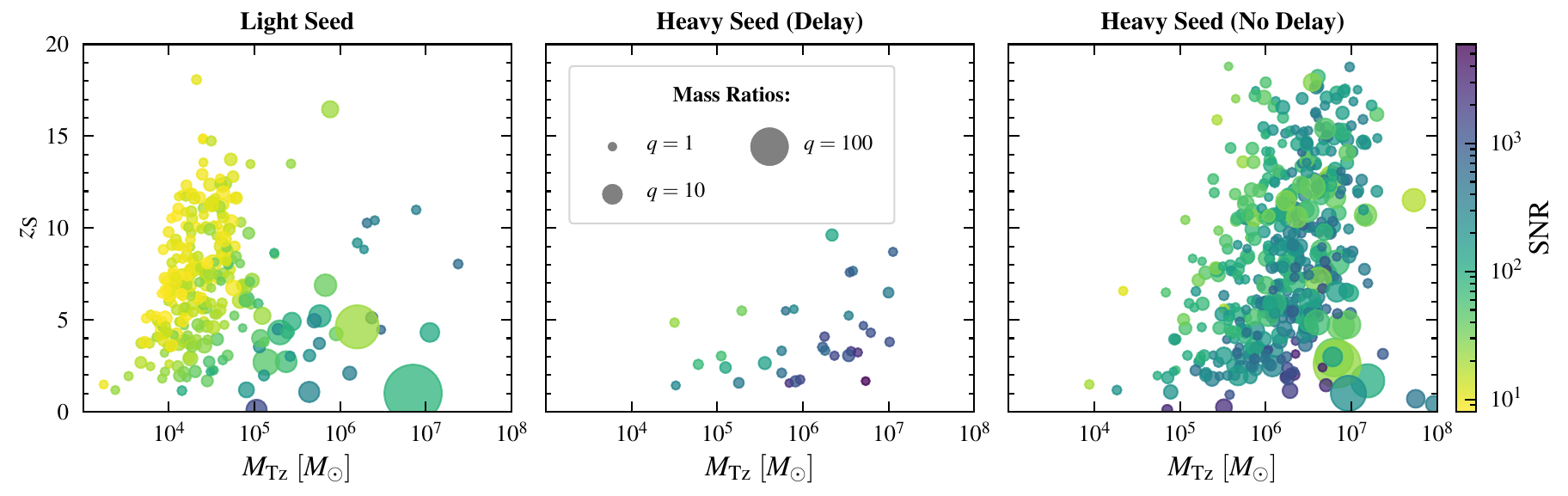}
    \caption{Comparison of the detectable MBHB population parameters for our three semi-analytical models: light seed, heavy seed (delay), and heavy seed (no delay).
    The scatter plots show the source redshift $z_{\rm S}$ as a function of the detector-frame total mass $M_{\rm Tz}$ for each MBHB in the given population.
    The area of each filled circle is proportional to $\sqrt{q}$, where $q$ is the mass ratio of the binary; some examples are shown in the legend, for reference.
    The color of the circles represents the SNR of the binary, as shown by the color bar on the right.
    The number of circles in each panel allows for an easy comparison of the number of observable MBHBs in the corresponding model.}
    \label{fig:source_pop}
\end{figure*}

\subsection{Model-Agnostic Approach}\label{sec:source_model_indep}

While there are multiple semi-analytical and simulation-based population models for MBHBs, the origin and evolution of these systems is poorly constrained~\cite{Berti:2006ew,Barack:2018yly,Volonteri:2021sfo}.
One of the main goals of the LISA mission is to understand the nature of massive black hole seeds, the combined effect of mergers and accretion on their growth, and the relation between MBHB mergers and structure formation~\cite{LISA:2022yao}.

Given the large uncertainties in the mass, redshift, and spin distributions of MBHBs, we first adopt a model-agnostic approach to better understand the lensing probability of different sources. To this end, we simulate $\orderOf(1000)$ MBHBs with a range of masses, redshifts, and mass ratios that broadly cover the parameter space of MBHBs observable by LISA. 

The main parameters characterizing the observability of an MBHB are its redshift $z_{\rm S}$, the detector-frame (redshifted) total mass $M_{\rm Tz}$, and the mass ratio $q$.
We sample this parameter space by considering 30 values for $M_{\rm Tz}$ spaced logarithmically in the range $[ 10^5, 10^8 ]\, \Msun$,
10 values of $z_{\rm S}$ in the range $[1, 10]$ chosen such that the sampling is uniform in comoving distance,
and 3 different values of the mass ratio: $q = 1, 5, 10$. 
The redshift is uniformly sampled in comoving distance because this is the quantity that directly influences the optical depth. For instance, the optical depth reaches its maximum when the lens is situated halfway between the observer and the source in terms of comoving distance.

As shown in Fig.~5 of Ref.~\cite{Caliskan:2022hbu}, the effect of aligned or antialigned spins is a relatively symmetrical correction of order unity on the measurement uncertainty of lens parameters. For simplicity and computational convenience, we set the spin magnitudes to zero: $\chi_1, \chi_2 = 0$.
For the other parameters of the binary, we choose values that, based on previous experience, lead to typical median measurement uncertainties: we fix the inclination angle to $\iota = \pi/3$, the polarization angle to $\psi = \pi/3$, the coalescence phase to $\phi_c = \pi/3$, the right ascension to $\lambda = \pi/3$, the declination to $\beta = \pi/3$, and the time of coalescence to $t_0 = 0$. As in Ref.~\cite{Caliskan:2022hbu}, we perform our analyses with the phenomenological waveform model \texttt{IMRPhenomHM}, which includes the full inspiral, merger, and ringdown of aligned-spin MBHBs as well as higher-order multipoles of the radiation~\cite{London:2017bcn}.

To accurately assess the optical depth of lensing (including WO effects) for a given MBHB, it is crucial to perform an extensive number of information-matrix calculations covering the relevant lens parameter space. 
This process is computationally intensive, requiring $\orderOf(10^4)$ calculations for each binary. 
We scan the parameters that most affect the lensing probability, i.e., the MBHB masses and redshift, which affect not only the SNR, but also the optical depth: see Eq.~\eqref{eq:optical_depth} below. 
In this way we can compute the lensing probabilities for $\sim 1000$ MBHBs.
A more detailed study of the effect of aligned spins, mass ratio, polarization and inclination angles, phase of coalescence, and sky location can be found in Ref.~\cite{Caliskan:2022hbu}.

\subsection{Semi-Analytical Population Models}\label{sec:source_pop_model}

The model-agnostic approach offers insights into lensing probabilities across a wide range of MBHB parameters. However, understanding existing population models can also be highly informative. In this regard, we explore three semi-analytical population models based on Ref.~\cite{Barausse:2012fy} (subsequently improved in Refs.~\cite{Sesana:2014bea,Antonini:2015cqa,Antonini:2015sza}). These models are frequently used in studies of various aspects of LISA science involving MBHBs, including event rates and parameter estimation~\cite{Klein:2015hvg}, the potential of MBHBs as standard sirens~\cite{Tamanini:2016zlh,LISACosmologyWorkingGroup:2019mwx,LISACosmologyWorkingGroup:2022jok}, and their applications in black hole spectroscopy and tests of general relativity~\cite{Berti:2016lat,Perkins:2020tra,Bhagwat:2021kwv}.

These population models employ a semi-analytical approach for galaxy formation within a $\mathrm{\Lambda CDM}$ universe, and they provide illustrative examples of the possible mass, redshift, and spin evolution of MBHBs. 
The models account for the effect of MBHBs on the development of structures, and they include the effect of nuclear galactic gas on the history of MBHB accretion and pre-merger spin-alignment~\cite{Barausse:2012fy, Klein:2015hvg}.
The evolution of MBHBs considers two seed scenarios (``light seeds'' and ``heavy seeds''). The light-seed scenario posits that the MBHB seeds originate from the remnants of Population III stars, which formed in low-metallicity environments within the $15 \lesssim z_{\rm S} \lesssim 20$ redshift range. This model includes delay times between the MBHB and galaxy mergers.
In the heavy-seed scenario, MBHBs form from the collapse of massive protogalactic disks within the $10 \lesssim z_{\rm S} \lesssim 15$ redshift range, and they already have masses of $\sim 10^5\ \Msun$ at high redshifts. 
To account for uncertainties in the solution of the so-called ``final parsec problem,'' we consider two heavy-seed models.\footnote{The ``final parsec problem'' refers to the theoretical challenge in astrophysics of identifying  physical mechanisms that can reduce the orbital separation of a MBHB by a factor of approximately 100, from an initial distance of about one parsec down to distances where gravitational radiation is the dominant source of dissipation, allowing the binary to merge within a Hubble time~\cite{Milosavljevic:2002ht,Merritt:2004gc}.} 
The ``heavy-seed (delay)'' model incorporates a model for the delay time between MBHB mergers and galaxy mergers, while the ``heavy-seed (no delay)'' model assumes that there is no delay. 
The light-seed and heavy-seed (delay) scenarios represent more realistic and conservative approaches, while the heavy-seed (no delay) scenario is the most optimistic: in this scenario the MBHBs are very massive, and their merger rates are highest since there is no time delay between the galaxy and MBHB mergers, resulting in higher number of events and mergers at potentially higher redshifts.

The masses, redshift, and time of coalescence of each binary in these three model populations are sampled assuming a 4-year LISA mission. The extrinsic binary parameters ($\iota, \psi, \phi_c, \lambda$ and $\beta$) are isotropically sampled in their respective ranges, and therefore, the effect of their variance on the lensing rate is accounted for. Finally, we set the dimensionless spin magnitudes $\chi_1=\chi_2=0$ for simplicity and due to great uncertainty in the spin distributions of MBHBs~\cite{Berti:2008af,Sesana:2014bea,Massonneau:2022uwg}. 
As in the model-agnostic approach of Sec.~\ref{sec:source_model_indep}, we use the $\texttt{IMRPhenomHM}$ waveform model~\cite{London:2017bcn}. We assume MBHB to be detected if their SNR is greater than a threshold $\rho_{\rm thr} = 8$.

The detectability of WO effects is highly sensitive to the source parameters, and in particular to $M_{\rm Tz}$ and $z_{\rm S}$, so it is important to understand the difference in the source parameters predicted by our three semi-analytical models.
These distributions are shown in Fig.~\ref{fig:source_pop}.
Each scatter plot displays the total redshifted mass $M_{\rm Tz}$ (horizontal axis) and the source redshift $z_{\rm S}$ (vertical axis) for each MBHB in the given population.
The size of each filled circle corresponds to the mass ratio of the binary, with larger circles denoting higher mass ratios. 
The color of the circles indicates the SNR of the binary. The SNR range spanned by the three populations can be seen from the color bar on the right.

The number of circles in each panel gives a visual sense of the number of MBHBs in the respective population. 
The number of detected binaries (those having $\rho>\rho_{\rm thr}$ over a ``fiducial'' 4-year LISA mission~\cite{Seoane:2021kkk}) is 282, 32, and 474 for the light-seed, heavy-seed (delay), and the heavy-seed (no delay) scenarios, respectively. Our rates closely match those in Ref.~\cite{Klein:2015hvg}, with minor differences due to our updated waveform model and to the use of a slightly different LISA noise power spectral density.

The heavy seed (no delay) population features the highest mass MBHBs, predominantly within the range $M_{\rm Tz} \in (10^6, 10^7)\, M_{\odot}$, followed by the heavy seed (delay) population, while detectable binaries in the light seed scenario mainly have $M_{\rm Tz}\in (10^4, 10^5)\, M_{\odot}$.
The heavy seed (no delay) population also produces the largest number of detectable high-redshift binaries: this observation is important in the context of lensing. While the heavy seed (delay) scenario produces a small number of events, most of these binaries have a mass ratio $q \sim 1$, and their masses are in the range where the WO optical depth peaks, so (as we will see below) this population still exhibits relatively high lensing probabilities.

\section{Lensing Probability}\label{sec:Probability}

The probability of lensing detection is intrinsically linked with the attributes of the lens and source, along with their respective populations. 
In the geometric optics limit, the cross-section for yielding multiple images depends on the lens properties and on the geometrical arrangement of the source-lens-observer system.
Typically, the critical impact parameter for SL is $y_{\rm cr}^{\rm SL} \sim 1$, with $y_{\rm cr}^{\rm SL} = 1$ being an exact result for SIS lenses.
Thus, the probability of having multiple images
is independent of the black hole binary source parameters in the SL scenario; an exception is the source redshift $z_{\rm S}$, which dictates the number of lenses encountered along the path. 
However, the probability of observing lensing within the SL regime, which takes into account the detection of these multiple images, might be reduced as some images can be demagnified (or insufficiently magnified). 
Consequently, the critical impact parameter can effectively become smaller, with a magnitude determined by the detector sensitivity.\footnote{A better source-lens-observer alignment (i.e., a lower impact parameter $y$) is required for a higher magnification ratio. For more details on the magnification ratios for strong lensing, using the same lens model adopted here, see Eq.~(A2) of Ref.~\cite{Caliskan:2022hbu}.} Even though the rate of SL is influenced by the properties of the black hole binary population, the optical depth associated with the occurrence of SL remains independent of the source parameters, with the exception of $z_{\rm S}$.

On the contrary, the WO optical depth is directly influenced by the source parameters~\cite{Caliskan:2022hbu}.
The measurement of WO effects hinges upon the identification of waveform distortions. The highest measurable impact parameter for WO effects $y_{\rm cr}^{\rm WO}$ (henceforth, for brevity, $y_{\rm cr}$) can be different from (and much larger than) unity. For instance, $y_{\rm cr}$ can extend to $\mathcal{O}(10-100)$~\cite{Caliskan:2022hbu}.
Moreover, $y_{\rm cr}$
is a function of the redshifted lens mass, decreasing as $M_{\rm Lz}$ increases. In addition, $y_{\rm cr}(M_{\rm Lz})$ can change by several orders of magnitude for MBHBs with different source parameters~\cite{Caliskan:2022hbu}.

\begin{figure*}[h!]
\includegraphics[width=\textwidth]{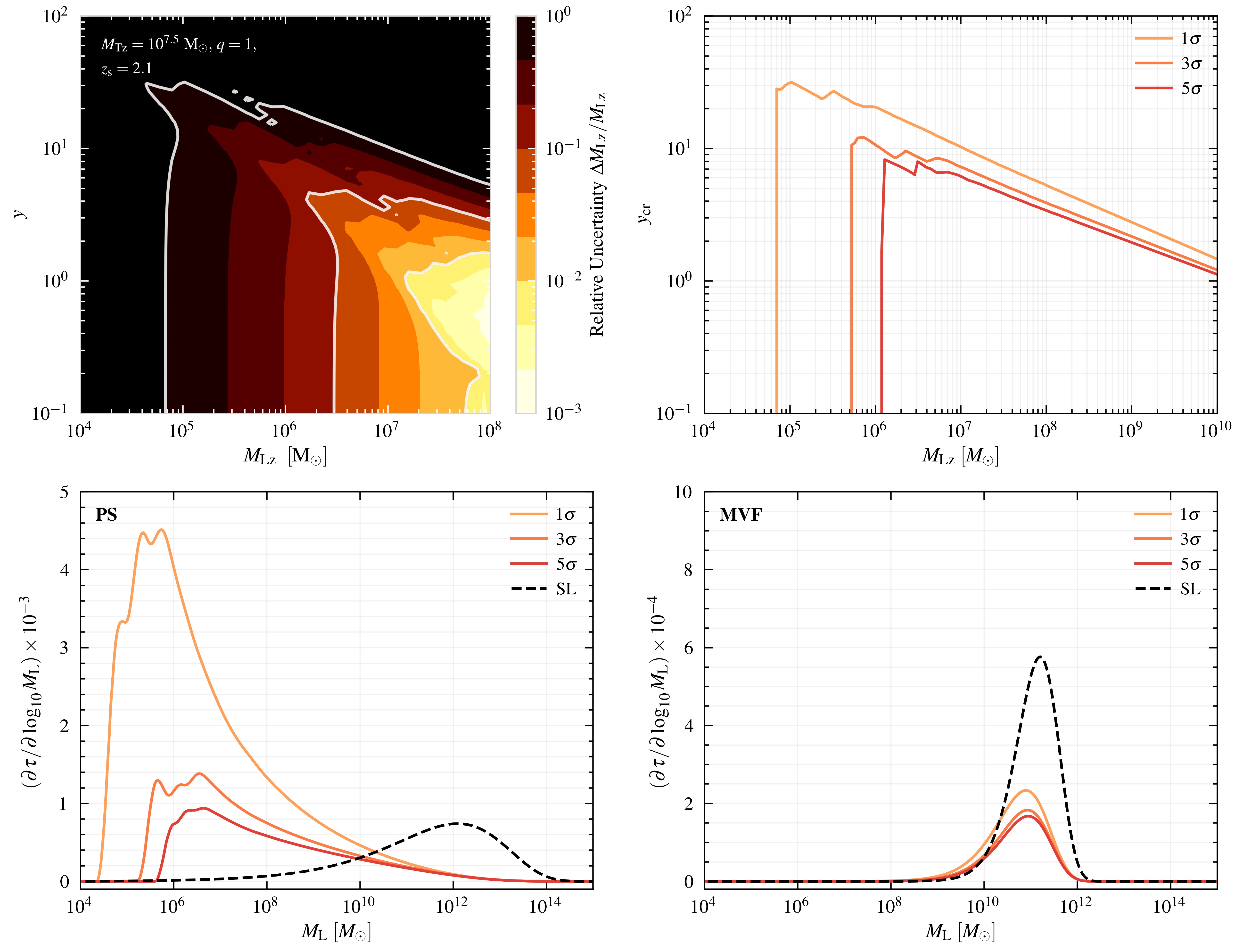} 
\caption{
Upper left panel: relative uncertainty of the redshifted lens mass measurement, $\Delta \lensMass/\lensMass$, in the $(\lensMass, y)$ plane, for an MBHB with the mass and redshift parameters listed in the legend. White contour lines represent $100\%$, $10\%$, and $1\%$ relative errors. In the black regions, relative errors exceed $100\%$.
Upper right panel: the critical (highest measurable) impact parameter, $y_{\rm cr}$, as a function of $\lensMass$. The findings correspond to three different detection thresholds based on the measurability of $\lensMass$ ($1\sigma$, $3\sigma$, and $5\sigma$), as determined in the upper left panel.
Lower left panel: the differential optical depth per $\log_{10}M_{\rm L}$ bin. 
The underlying lens population is based on the PS halo mass function. 
The colored lines exhibit the differential optical depth for wave-optics (WO) effects under three different detection thresholds for this MBHB. 
The dashed black line denotes the maximum differential optical depth for strong lensing (SL). 
This line presupposes the critical impact parameter of SL is $y_{\rm cr}^{\rm SL} = 1$, and it does not account for any false alarms or images potentially below the detection threshold. 
Consequently, the SL (differential) optical depth can be viewed as a theoretical upper limit. 
In some regions, the SL optical depth is overtaken by the WO optical depth. Here, the lens parameters are such that the geometric optics approximation is invalid, making WO the only observable feature. 
The WO optical depths peak for lower $M_{\rm L}$, indicating that LISA can constrain the number density of low-mass lenses (such as dark matter substructures). 
Lower right panel: same as the lower left panel, but for an underlying lens population based on the measured velocity function (MVF).
The WO optical depths are considerably reduced in this scenario, and often smaller than the SL optical depths (note the order-of-magnitude difference in the vertical labels). 
The bottom panels suggest that LISA might fail to detect any WO effects if the abundance of low-mass lenses is significantly lower than that predicted by the PS halo mass function.
}
\label{fig:2by2}
\end{figure*}

The value of $y_{\rm cr}(M_{\rm Lz})$ depends on the SNR, with higher SNRs leading to larger values of $y_{\rm cr}(M_{\rm Lz})$, and it is also sensitive to the inclusion of higher-order multipoles in the MBHB waveform model.
This is because WO effects are \textit{frequency-dependent}, and the interplay between detector sensitivity and frequency evolution (including higher-order modes of the radiation) determines their observability.
Higher-order modes also help because they break the degeneracy between different binary parameters.

In this section, we elucidate the methodology to compute the probability of observing lensing with measurable WO signatures as a function of the lens and source parameters.
We first define the optical depth $\tau(\bm \theta^{\rm S})$, where $\bm \theta^{\rm S}$ represents the list of source parameters. 
We then describe how the probability can be calculated for a specific MBHB using various detection thresholds. 
Finally, we examine how assumptions on the underlying lens population influence the optical depth.

\subsection{Optical Depth} \label{sec:optical_depth_calculation}

The optical depth $\tau (\bm \theta^{\rm S})$ for an MBHB with source parameters $\bm\theta^{\rm S}$ at redshift $z_{\rm S}$ is given by~\cite{Schneider:1992, Takahashi:2003ix}:
\begin{widetext}
\begin{equation} \label{eq:optical_depth}
    \tau (\bm \theta^{\rm S} = \{z_{\rm S}, \dots \}) = \int_{0}^{z_{\rm S}} \mathrm{d}z_{\rm L} \int_{M_{\rm L}^{\rm min}}^{M_{\rm L}^{\rm max}} \mathrm{d}M_{\rm L}\, \frac{4 \pi M_{\rm L} D_{\rm LS}}{D_{\rm L} D_{\rm S}}\, y^2_{\rm cr}(M_{\rm L}|\bm \theta^{\rm S})\, n\left[M_{200}, z_{\rm L}, z_{\rm S}\right]\, \chi^2(z_{\rm L}) \frac{\mathrm{d}\chi(z_{\rm L})}{\mathrm{d}z_{\rm L}} \frac{\mathrm{d}M_{200}}{\mathrm{d}M_{\rm L}}\,.
\end{equation}
\end{widetext}
The parameters appearing in this equation are the lens redshift $z_{\rm L}$, the lens mass $M_{\rm L}$, the critical (maximum measurable) impact parameter $y_{\rm cr}$, the comoving number density of halos $n$, the comoving distance at a given redshift $\chi$, and the halo virial mass $M_{200}$, 
respectively. 
Note that, within the integrand, $M_{200}$ is a function of $M_{\rm L}$, $z_{\rm L}$, and $z_{\rm S}$, as elucidated in Sec.~\ref{sec:mass_function_to_profile}.

The $z_{\rm L}$ integral ranges from the observer (at redshift zero) to the source (at redshift $z_{\rm S}$).
The $M_{\rm L}$ integral ranges between the minimum and maximum measurable lens masses ($M_{\rm L}^{\rm min}$ and $M_{\rm L}^{\rm max}$, respectively), which in turn depend on $\bm \theta^{\rm S}$.
The parameter $y_{\rm cr}$ is also a function of $\bm \theta^{\rm S}$ and $M_{\rm L}$. In fact, we first compute $y_{\rm cr}$ (which depends on the chosen confidence level for detection, as we explain below) and then define the lens mass integration limits, $M_{\rm L}^{\rm min}$ and $M_{\rm L}^{\rm max}$, as the points where $y_{\rm cr}$ approaches zero.

The optical depth computed via Eq.~\eqref{eq:optical_depth} correlates with the lensing observation likelihood. The lensing probability for an MBHB at source redshift $z_{\rm S}$ is expressed as:
\begin{equation} \label{eq: probability_of_lensing}
P = 1 - e^{-\tau(\bm \theta^{\rm S} = \{z_{\rm S}, \dots \})}\,.
\end{equation}
For low-optical-depth scenarios, we have $P\simeq \tau$.

\subsection{Calculating the Probability} \label{sec:calculating_the_probability}

To compute the probability of observing WO effects, we first need to determine the critical impact parameter $y_{\rm cr}(M_{\rm L})$, which in turn depends on the measurability of lens parameters for a given MBHB.

For illustration, in Fig.~\ref{fig:2by2} (upper left panel) we show the relative uncertainty in the redshifted lens mass measurement $\Delta M_{\rm Lz} / M_{\rm Lz}$ as a function of $M_{\rm Lz}$ and of the impact parameter $y$ for an MBHB with $M_{\rm Tz} = 10^{7.5}\ M_\odot$, $q=1$, and $z_{\rm S} = 2.1$. 
White contour lines mark $100\%$, $10\%$, and $1\%$ relative uncertainties. The results shown in this plot are in good agreement with the information-matrix calculations of Ref.~\cite{Caliskan:2022hbu} but the computational time is significantly shorter because we now use the lookup table described in Appendix~\ref{App:LUT} rather than analytical solutions for $F(w, y)$ and its derivatives.

Color gradients indicate measurement error levels, with black regions signifying relative uncertainties exceeding $100\%$.
If $(M_{\rm Lz}, y)$ coincides with a point within the black region, this indicates that the $1\sigma$ confidence interval (CI) of the posterior
for $M_{\rm Lz}$ encompasses $M_{\rm Lz} = 0$, i.e., it is compatible with the absence of lensing within $1\sigma$.

The choice for $y_{\rm cr}(M_{\rm Lz})$ is somewhat subjective. For example, we could set a threshold at the $100\%$ relative uncertainty contour line: this would ensure that for any $(M_{\rm Lz}, y)$ within the region, the $M_{\rm Lz}$ posterior is consistent with lensing at $1\sigma$ or higher.
This choice would not be conservative: lens parameters along this contour line still yield a posterior that is consistent with the absence of lensing for CIs exceeding $1\sigma$. 
More conservative choices would require the lens parameter posteriors to be consistent with lensing within the $3\sigma$ (or $5\sigma$) CI. Then our $y_{\rm cr}(M_{\rm Lz})$ would be based on the contour line where the relative uncertainty is $1/3$ (or $1/5$). 

Note that $y_{\rm cr}$ is not constant: it decreases monotonically with $M_{\rm Lz}$. 
This dependency must be taken into account when we compute the optical depth, and it can be understood as follows.
In the top left panel of Fig.~\ref{fig:2by2}, the lens parameters are determined through the frequency-dependent WO effects.
The geometric optics approximation holds when $f t_{\rm d} \gg1$, where $f$ is the GW frequency and $t_{\rm d}$ is the lensing-induced time delay.
In our lens model, the condition reads $wy \propto M_{\rm Lz} f y \gg 1$.
This shows that the geometric optics approximation should be valid at lower $y$ as $M_{\rm Lz}$ increases, and it is consistent with the observed trend between $y_{\rm cr}$ and $M_{\rm Lz}$: when the geometric optics approximation is valid, instead of observing multiple waveforms for all $y>1$, we would observe a single waveform, leading to a degeneracy of the lens and source parameters. Conversely, for a lens model that still produces multiple images when $y > 1$ (such as a point-mass lens) $y_{\rm cr}$ will not decrease with $M_{\rm Lz}$, but it will remain constant: see Fig.~7 of Ref.~\cite{Caliskan:2022hbu}.

In the top left panel of Fig.~\ref{fig:2by2} we show $\Delta M_{\rm Lz} / M_{\rm Lz}$ and use it to determine $y_{\rm cr}$.
We could have enforced a similar criterion based on the impact parameter measurement, $\Delta y/y$,
but we chose not to do so for the following reasons.
First of all, the trend of $y_{\rm cr}$ deduced using $\Delta y/y$ closely resembles that obtained through $\Delta M_{\rm Lz} / M_{\rm Lz}$ across most of the lens parameter space.\footnote{Several examples of contour plots for the measurement uncertainty of the impact parameter, $(\Delta y/y$), can be found in Fig.~8 of Ref.~\cite{Caliskan:2022hbu}.} Notable differences arise in the low impact parameter range, where $M_{\rm Lz}$ can still be measured, but $\Delta y/y$ remains high. This, however, does not imply an inconsistency with lensing.
The second reason is precisely that, unlike having a measurement of the redshifted lens mass consistent with zero, $y \sim 0$ does not indicate no lensing: it only means that the alignment between the source-lens-observer system is near perfect.
Therefore, $\Delta y/y > 1$ 
does not necessarily rule out lensing, while a posterior of the lens mass consistent with zero does.

From now on, we will consider three detection thresholds ($1\sigma$, $3\sigma$, and $5\sigma$) to compute $y_{\rm cr}$ for a given MBHB using the procedure described above. 
These detection thresholds identify the range of lens parameters for which the posteriors are consistent with lensing (with \textit{observable} WO effects) within the corresponding CIs.

In the upper right panel of Fig.~\ref{fig:2by2}, we show fits of $y_{\rm cr}(M_{\rm Lz})$ (specific to the given MBHB) for the three different detection thresholds (in color).
These fits can be used to compute the optical depth through Eq.\eqref{eq:optical_depth}.
Note that $y_{\rm cr}$ is a function of $M_{\rm Lz}$, rather than $M_{\rm L}$. For optical depth calculations, we account for this nuance and find $y_{\rm cr}$ as a function of $M_{\rm L}$ for each $z_{\rm L}$ in the integration.

In the lower-left panel of Fig.~\ref{fig:2by2}, we display the differential optical depth per $\log_{10}$ lens mass bin, $\partial \tau / \partial \log_{10}{M_{\rm L}}$, focusing once again on this particular MBHB. In the calculation of the comoving number density $n$ in Eq.~\eqref{eq:optical_depth} we assume that the lens population follows the PS halo mass function (refer to Sec.~\ref{sec:PS_halo_mass_function}). 
As such, this lens population incorporates all virialized isolated halos.
The colored lines indicate the differential optical depth for WO effects at three detection thresholds, while the dashed black line represents the maximum differential optical depth for SL. 
This dashed line (which assumes a critical impact parameter for SL $y_{\rm cr}^{\rm SL} = 1$, and does not account for potential false alarms or detection thresholds) serves as a theoretical upper bound.

The differential optical depth for SL is smaller than the WO optical depth at low lens masses ($M_{\rm L} \lesssim 10^{10}\ M_\odot$), where the key assumption underlying SL (i.e., the geometric optics approximation) is invalid. Hence, only WO effects are measurable in these regions. We still show the SL differential optical depth in this region for comparison, although it is actually zero since multiple images do not form in the WO regime.

The differential optical depth for SL peaks at high values of $M_{\rm L}$ (comparable to galaxy masses), while the differential optical depth for WO has a maximum at significantly lower masses. This indicates that most of the WO detection probability originates from low-mass lenses, including DM substructures. This is an important conclusion of our work: LISA can efficiently constrain the number density of these elusive low-mass DM halos and subhalos, which are otherwise challenging to investigate, since they are non-luminous and have negligible SL optical depths (because of their small Einstein radii). Depending on the detection rate and confidence levels, LISA can also restrict these structures' density profiles. 
Regardless of the chosen detection threshold, the overall optical depth for WO is either substantially higher than (or comparable to) that of SL. This suggests that past LISA lensing studies omitting WO effects underestimated the detection probability.

\begin{figure*}[t]
    \centering
    \includegraphics[width=.99\textwidth]{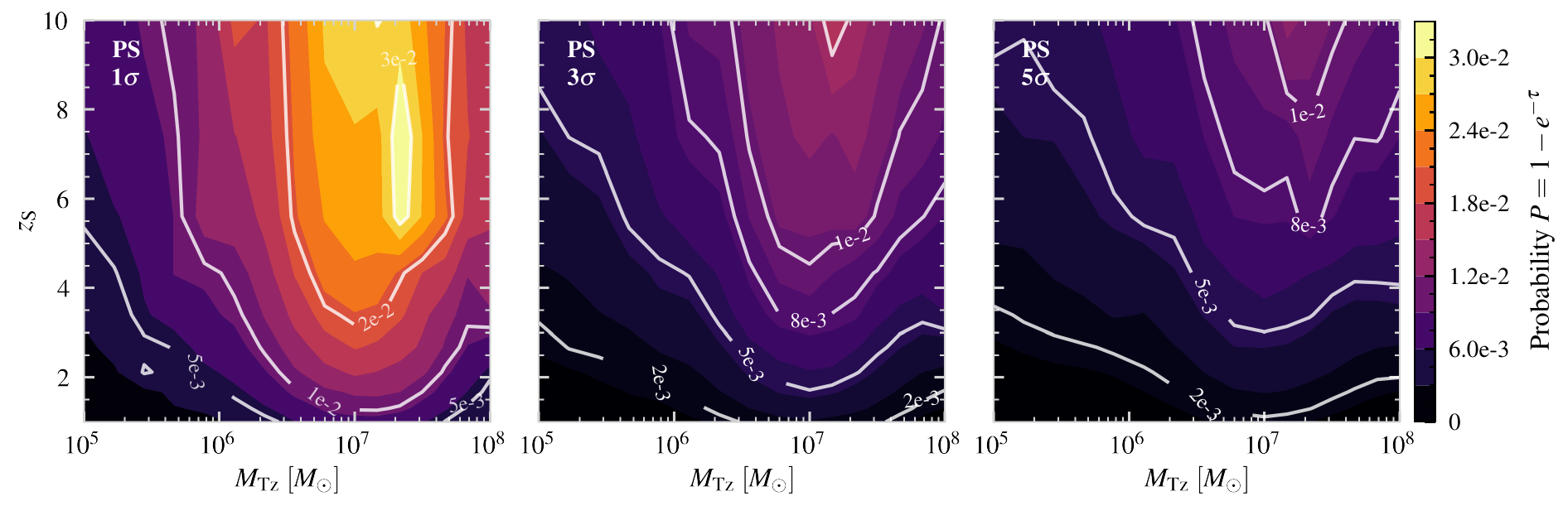}
    \caption{The probability of observing WO effects assuming a lens population based on the PS halo mass function. 
    The results are given for various confidence levels ($1\sigma$, $3\sigma$, and $5\sigma$) and for equal-mass, nonspinning binaries with varying source parameters: detector-frame (redshifted) source total mass $M_{\rm Tz} \in [10^5, 10^8]\ \Msun$ and source redshift $z_{\rm S} \in [1, 10]$. 
    The probability can be as large as $\sim 3\%$ and it peaks around $M_{\rm Tz} \approx 10^7$. 
    The probability rises with $z_{\rm S}$, but this is compensated by the decline in $y_{\rm cr}$ due to the low SNRs of high-redshift sources.
    }%
    \label{fig:optical_depth_PS}%
\end{figure*}

The lower right panel of Fig.~\ref{fig:2by2} is similar to the lower left panel, but now we compute $n$ using the MVF lens population.
As we discussed in Sec.~\ref{sec:Measured_velocity_function}, this observation-based lens population can be considered as a conservative (but robust) lower boundary.
Compared to the PS model, the MVF model leads to considerably reduced optical depths: note the order-of-magnitude difference in the vertical axis between the two plots. 

The discrepancy stems from several factors. The detectability of WO effects mostly comes from the fact that $y_{\rm cr}$ can be large for low-mass lenses. The differential optical depth scales with $y_{\rm cr}^2$, and therefore the WO optical depth is significantly enhanced when there is an abundance of low-mass lenses. These lenses, with their minuscule Einstein radii, barely contribute to the SL optical depths.
The MVF model has a dearth of low-mass lenses compared to the PS model, resulting in a comparatively low number density where $y_{\rm cr}$ is highest. For this reason, the MVF differential optical depth is broader and peaks at larger lens masses. Regrettably, the observational uncertainties translate into significant uncertainties in the overall probability of observing WO effects: the MVF-based lower bound is roughly an order of magnitude smaller than the PS-based estimate. 

The results in Fig.~\ref{fig:2by2} refer to a specific MBHB, but our qualitative considerations apply to any MBHB detectable by LISA. In the following section, we will focus on the MBHB populations predicted in three specific formation scenarios. This will allow us to better quantify the probability of observing WO effects, as well as the lens mass and redshift ranges probed by LISA.

\section{Results} \label{sec:Results}

In this section, we compute the likelihood of LISA detecting WO effects in GWs from MBHBs. In Sec.~\ref{sec:results_model_agnostic}, we adopt a model-agnostic approach, in which we simply scan the source mass, mass ratio, and redshift ranges observable by LISA (see Sec.~\ref{sec:source_model_indep}). In Sec.~\ref{sec:results_pop_model}, we compute the lensing rates for the three semi-analytical MBHB population models described in Sec.~\ref{sec:source_pop_model}.

As anticipated in Sec.~\ref{sec:Lens_Population}, we use two lens populations. The first is theory-based, underpinned by the PS halo mass function (incorporating all virialized isolated halos). The second is observation-based, drawn from the MVF. The probabilities are computed following the method described in Sec.~\ref{sec:Probability}. Following this analysis, we delve into the ranges of halo mass (Sec.~\ref{sec:results_lens_mass}) and redshift (Sec.~\ref{sec:results_lens_redshift}) that are accessible by observing WO effects.

\begin{figure*}[t]
    \centering
    \includegraphics[width=.99\textwidth]{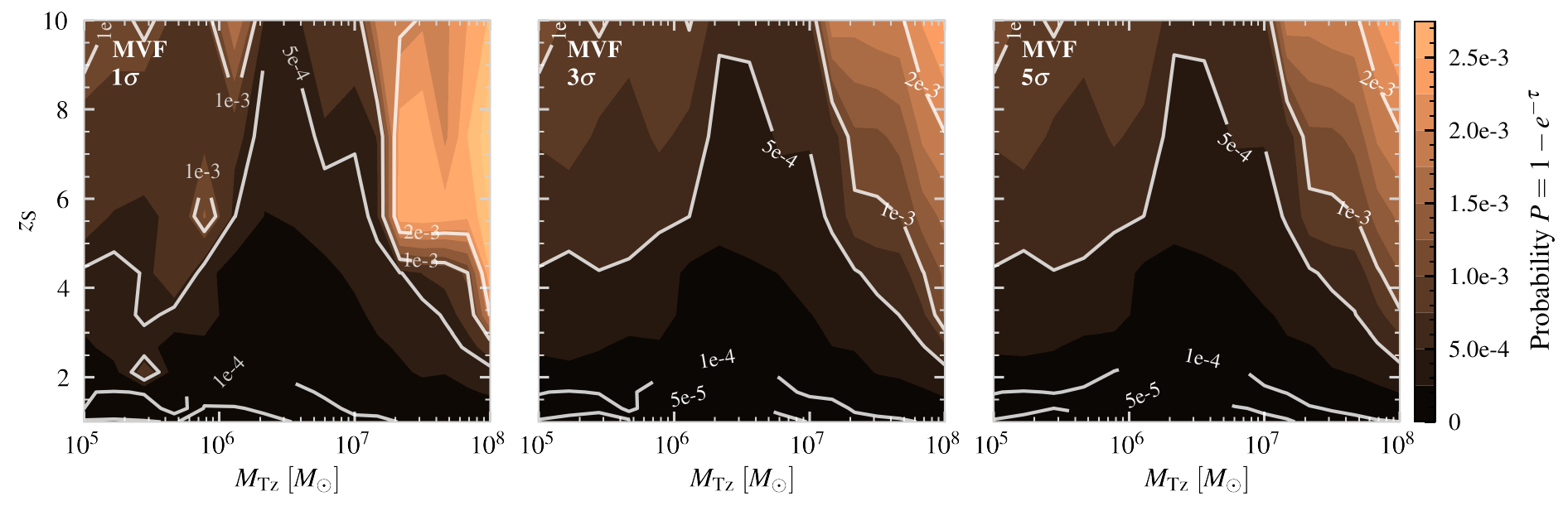}
    \caption{Same as Fig.~\ref{fig:optical_depth_PS}, but with the lens population predicated on the MVF derived from \texttt{SDSS'06}~\cite{Choi:2006qg, SDSS:2007aih}, which serves as a conservative lower-bound estimate for the lens population. 
    Optical depth values across the parameter space are consistently one order of magnitude lower, so it is possible to have no events with observable WO effects during the LISA mission.
    }
    \label{fig:optical_depth_MVF}
\end{figure*}

\subsection{Probabilities for the Model-Agnostic Approach} \label{sec:results_model_agnostic}

\subsubsection{Press-Schechter Halo Mass Function}\label{sec:results_mod_indep_PS}

In Fig.~\ref{fig:optical_depth_PS}, we focus on equal mass ($q=1$), non-spinning MBHBs and plot the probability of observing WO effects as a function of the other two intrinsic source parameters: the detector-frame (redshifted) total mass $M_{\rm Tz} \in [10^5, 10^8]\, \Msun$, and the redshift $z_{\rm S} \in [1, 10]$. Here, the lens population is based on the PS halo mass function, which includes all virialized isolated halos. We provide the results for the detection confidence levels $1\sigma$, $3\sigma$, and $5\sigma$ in the left, middle, and right panels, respectively.

For the range of sources considered, the probabilities are roughly between $0.5\%\, \textrm{--}\, 3\%$, $0.2\%\, \textrm{--}\, 2.0\%$, and $0.2\%\, \textrm{--}\, 1.0\%$ for $1\sigma$, $3\sigma$, and $5\sigma$ thresholds, respectively. 
With a $1\sigma$ threshold, MBHBs with $M_{\rm Tz} \sim 2\times 10^7\ M_{\odot}$ and $z_{\rm S} \sim 5\, \textrm{--}\, 9$ yield the highest probabilities, highlighting that MBHBs with source-frame total masses $M_{\rm T}\sim 2\, \textrm{--}\, 3 \times 10^6\ M_{\odot}$ are optimal candidates for lensing. Similar findings emerge at stricter detection thresholds.

The existence of an optimal $M_{\rm Tz}$ is the result of two competing effects. As $M_{\rm Tz}$ increases, the signal wavelength gets larger, enhancing the likelihood of observing WO effects in higher-mass lenses. 
However, increasing $M_{\rm Tz}$ also leads to binaries getting \textit{redshifted} out of band, causing a significant reduction in their SNRs. This in turn shrinks the range of lens parameters for which WO effects can be detected, decreasing the lensing probability.

Sources with smaller $M_{\rm Tz}$ emit GWs at smaller wavelengths, potentially increasing the sensitivity to low-mass lenses with a high comoving number density. However, these low-mass binaries have low SNR at large source redshifts $z_{\rm S}$, leading to reduced probabilities. Low-redshift sources have higher SNR, but also lower optical depth.

Let us now turn to the redshift dependence. Lensing probabilities generally increase with $z_{\rm S}$. 
However, this increase is not as rapid for WO effects as it is for SL.
High $z_{\rm S}$ means more intervening lenses, increasing the optical depth, but it also means lower SNR, so only a subset of lens parameters yield observable WO effects (cf.~Fig.~\ref{fig:2by2}). 

In essence, the probability for a given binary is determined by the interplay between the comoving number density of lenses, the critical impact parameter $y_{\rm cr} (M_{\rm L})$, the signal wavelength, the presence of higher-order modes, and LISA's power spectral density. We find that the highest probabilities occur for high-SNR sources at $z_{\rm S} \gtrsim 4$.

Results for binaries with $q=5$ and $q=10$ are given in Appendix~\ref{App:Higher_Mass_Ratios}. In general, higher mass ratios result in probabilities that are about $30\%$ to $50\%$ lower compared to equal-mass binaries.

\begin{table*}[t]
\renewcommand{\arraystretch}{1.5}
\caption{Wave-optics lensing rates for the three MBHB population models of Sec.~\ref{sec:Source_Population} and for our two lens populations (PS and MVF). 
We list the number of detected events $(N_{\rm detect})$, the number of lensed events assuming different detection thresholds $(N_{\rm lensed}^{1\sigma, 3\sigma, 5\sigma})$, and the resulting lensing rates for each threshold.
The results refer to a ``fiducial'' 4-year LISA mission~\cite{Seoane:2021kkk}. Entries where the expected number of lensed events is larger than 1 are marked in boldface.}
\begin{ruledtabular}
\begin{tabular}{llccccc}
Source Population &
  Lens Population &
  $N_{\rm detect}$ &
  $N_{\rm lensed}^{1\sigma}$ &
  $N_{\rm lensed}^{3\sigma}$ &
  $N_{\rm lensed}^{5\sigma}$ &
  Lensing Rate $\{1\sigma, 3\sigma, 5\sigma\}\ [\%]$ \\
\colrule
Heavy Seed (No Delay) & PS  & 474 & \textbf{7.96} & \textbf{4.13} & \textbf{3.24} & $\{1.68,\ 0.87,\ 0.68\}$ \\
Heavy Seed (No Delay) & MVF & 474 & 0.42 & 0.36 & 0.35 & $\{0.09,\ 0.07,\ 0.07\}$ \\
Heavy Seed (Delay)    & PS  & 32  & 0.47 & 0.21 & 0.15 & $\{1.47,\ 0.65,\ 0.47\}$ \\
Heavy Seed (Delay)    & MVF & 32  & 0.01 & 0.009 & 0.008 & $\{0.03,\ 0.03,\ 0.03\}$ \\
Light Seed            & PS  & 282 & \textbf{1.51} & 0.57 & 0.37 & $\{0.53,\ 0.20,\ 0.13\}$ \\
Light Seed            & MVF & 282 & 0.02 & 0.01 & 0.01 & $\{0.007,\ 0.004,\ 0.004\}$
\end{tabular}
\end{ruledtabular}
\label{tab:population_results}
\end{table*}

\subsubsection{Measured Velocity Function}\label{sec:results_mod_indep_MVF}

Figure~\ref{fig:optical_depth_MVF} is similar to Fig.~\ref{fig:optical_depth_PS}, with the only difference being the assumed underlying lens population drawn from \texttt{SDSS'06}~\cite{SDSS:2007aih} MVF.
The probability peaks at approximately $0.3\%$, $0.2\%$, and $0.2\%$ for the $1\sigma$, $3\sigma$, and $5\sigma$ thresholds, respectively. Unlike the PS case, the highest probabilities correspond to the heaviest MBHBs. This is because, in the MVF case, the majority of the SL optical depth corresponds to higher lens masses
(see the discussion in Sec.~\ref{sec:Probability}).
Note also that the variation in probabilities for different detection thresholds is not as pronounced as in the PS case. This is because $y_{\rm cr}$ does not change much with the detection thresholds when $M_{\rm Lz}$ is high: see the upper right panel of Fig.~\ref{fig:2by2}. Given that most of the optical depth for the MVF case comes from high-mass lenses, different detection thresholds can be expected to yield similar (and low) results.

Overall, MVF-based probabilities are generally at least one order of magnitude lower than PS-based probabilities. Given the expected merger rate of MBHBs, an MVF-like lens distribution is likely to result in no events with observable WO effects.
This is only a conservative, observation-based lower bound.

\subsection{Probabilities for the Semi-Analytical Population Models} \label{sec:results_pop_model}

We now compute the WO event rates predicted by the population models of Sec.~\ref{sec:source_pop_model}. As a reminder, we employ the \texttt{IMRPhenomHM} waveform model, use the full LISA detector response, and account for any potential degeneracies among the lens and source parameters.

Table~\ref{tab:population_results} lists the lensing rates for the three MBHB populations and for the two lens population models (PS and MVF).
The best-case scenario corresponds to the heavy seed (no delay) model combined with the PS lens population: then the anticipated number of lensed events is $\sim 8$, $\sim 4$, and $\sim 3$ for the $1\sigma$, $3\sigma$, and $5\sigma$ detection thresholds, respectively. The light seed scenario with the PS lens population is the only other case with at least one expected event (at the $1\sigma$ threshold) during the LISA mission.

The WO lensing rates for the PS lens population are significant. 
For both heavy seed scenarios, the rates are in the range $0.5\%\, \textrm{--}\, 1.5\%$ for all detection thresholds. For comparison, this rate is larger than the SL rates of quasars~\cite{Treu:2005aw, Saha:2006cx, 2012AJ....143..119I} and supernovae~\cite{Kelly:2014mwa, 2018NatAs...2..324R, Goobar:2016uuf}, and larger than the anticipated lensing rate for stellar-mass black hole binaries~\cite{Xu:2021bfn, LIGOScientific:2023bwz}.

\begin{figure*}[t]
    \centering
    \includegraphics[width=.99\textwidth]{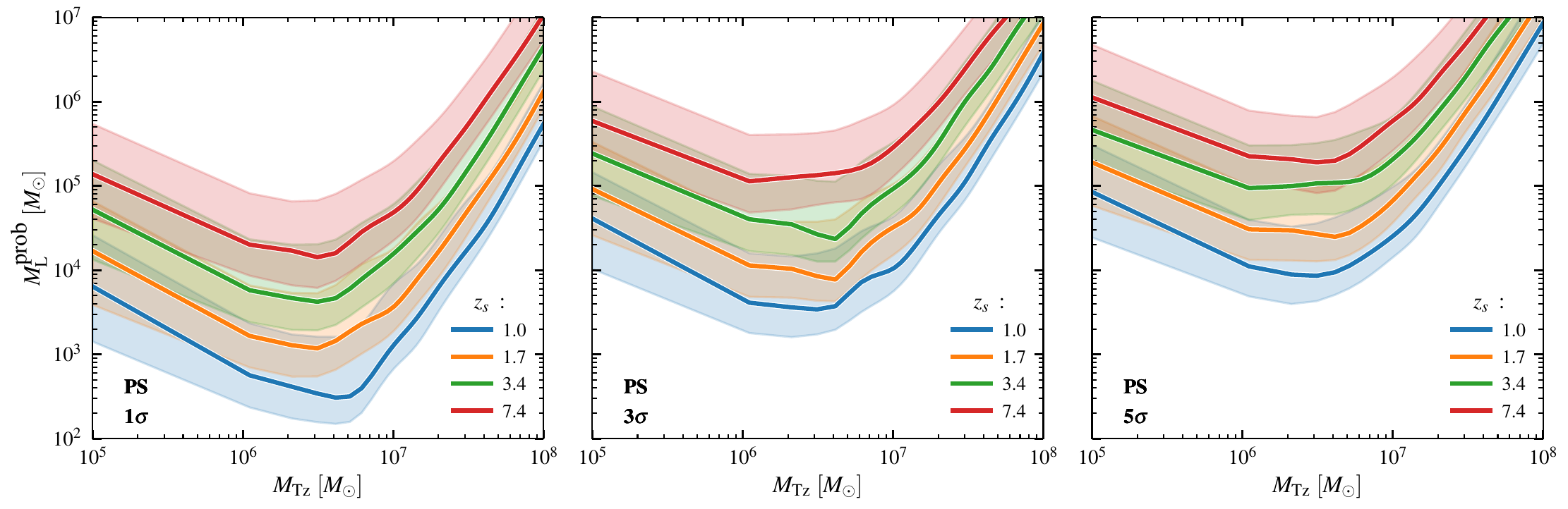}
    \caption{Median and $68\%$ confidence region (high-opacity and low-opacity areas, respectively) of the probed lens mass $M_{\rm L}^{\rm prob}$ as a function of $M_{\rm Tz}$.
    We consider the PS halo mass function, four selected source redshifts $z_{\rm S}$, and three detection confidence levels ($1\sigma$, $3\sigma$, and $5\sigma$).
    LISA can probe smaller DM halos compared to current observations of SL~\cite{Hezaveh:2016, Sengul:2021lxe, Sengul:2023olf}.
    }%
    \label{fig:M_L_inferred_PS_allSigma}%
\end{figure*}

\begin{figure*}[t]
    \centering
    \includegraphics[width=.99\textwidth]{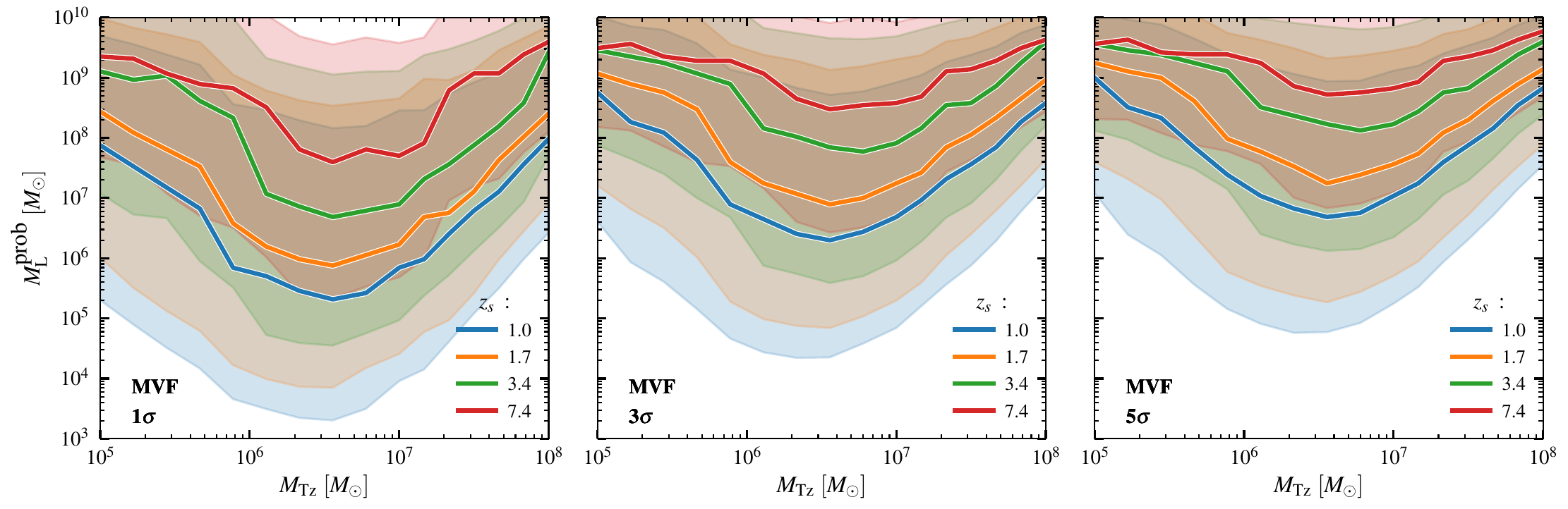}
    \caption{Same as in Fig.~\ref{fig:M_L_inferred_PS_allSigma}, but for the MVF lens population. 
    The inferred lens masses are significantly larger because the MVF WO optical depth has support at higher lens masses.
    }%
    \label{fig:M_L_inferred_MVF_allSigma}%
\end{figure*}

Another trend is clear: leaving aside the total number of events (which is very uncertain), the heavy seed (no delay) MBHB population yields the highest WO lensing rates, followed by the heavy seed (delay) and light seed scenarios.
This is because (as shown in Fig.~\ref{fig:source_pop}) binaries in the heavy seed (no delay) scenario have the highest SNR and source redshift. Moreover, the distribution of $M_{\rm Tz}$ is concentrated around higher masses, which further increases the lensing probability: compare Figs.~\ref{fig:optical_depth_PS},~\ref{fig:optical_depth_PS_q5}, and~\ref{fig:optical_depth_PS_q10}.
The heavy seed (delay) scenario yields slightly lower rates because the delay time causes binaries to merge at lower redshift, decreasing the optical depth.
In addition, the delay time leads to over an order of magnitude reduction in detected binaries. These two effects combined account for the drastic contrast in the number of lensed events $N_{\rm lensed}$, despite the similarity in the lensing rates.

The rates in the light seed scenario are approximately three times smaller than in the heavy seed scenarios for two reasons: the SNRs are usually lower, and so is the typical value of $M_{\rm Tz}$. Both effects are detrimental to the WO optical depth. The number of lensed events for the light seed population still outnumbers the heavy seed (delay) scenario, thanks to the substantial difference in the overall number of detections.

For the MVF case, both $N_{\rm lensed}$ and the lensing rates are disappointingly low and compatible with no WO observations over the fiducial duration of the LISA mission.

In conclusion, Table~\ref{tab:population_results} is consistent with expectations based on the properties of the MBHB populations (Fig.~\ref{fig:source_pop}) and on the lensing probabilities (Figs.~\ref{fig:optical_depth_PS},~\ref{fig:optical_depth_MVF},~\ref{fig:optical_depth_PS_q5}, and~\ref{fig:optical_depth_PS_q10}).

\subsection{Halo Mass Range Probed by LISA} \label{sec:results_lens_mass}

From a cosmological point of view, it is important to understand LISA's potential to probe the nature of DM. 
In this subsection and the next we ask: what is the range of halo masses and redshifts accessible to LISA through the observation of WO effects?
As far as we know, these questions (detectable range of halo masses and redshifts) were not addressed before.

We first apply the methods of Sec.~\ref{sec:Probability} to identify the range of lens masses that can be probed with LISA. We begin by computing the differential optical depth per lens mass bin (lower panels of Fig.~\ref{fig:2by2}) to determine the probability distribution of lens masses capable of lensing for a given MBHB. Once the differential optical depth is known, the mode, median, or a specified confidence interval of the distribution can be used to determine the most probable lens masses and redshifts.

We first discuss the PS lens population, and then the MVF lens population.

\subsubsection{Press-Schechter Halo Mass Function}

In Fig.~\ref{fig:M_L_inferred_PS_allSigma}, we plot the probed lens mass\footnote{In Fig.~\ref{fig:Mvir_Mlens} of Appendix~\ref{App:lens_mass_vs_halo_mass} we show, for reference, the correlation between lens mass and halo (virial) mass for several source and lens redshift configurations.} 
(denoted by $M_{\rm L}^{\rm prob}$) as a function of $M_{\rm Tz}$ for equal-mass binaries, assuming the PS-based lens population. 
The high opacity line and the low opacity region show the lens masses corresponding to the median and the 68\% confidence region of the differential optical depth (per lens-mass bin), respectively. The colors correspond to different source redshifts, while the three panels refer to different detection thresholds.

\begin{figure*}[t]
    \centering
    \includegraphics[width=.99\textwidth]{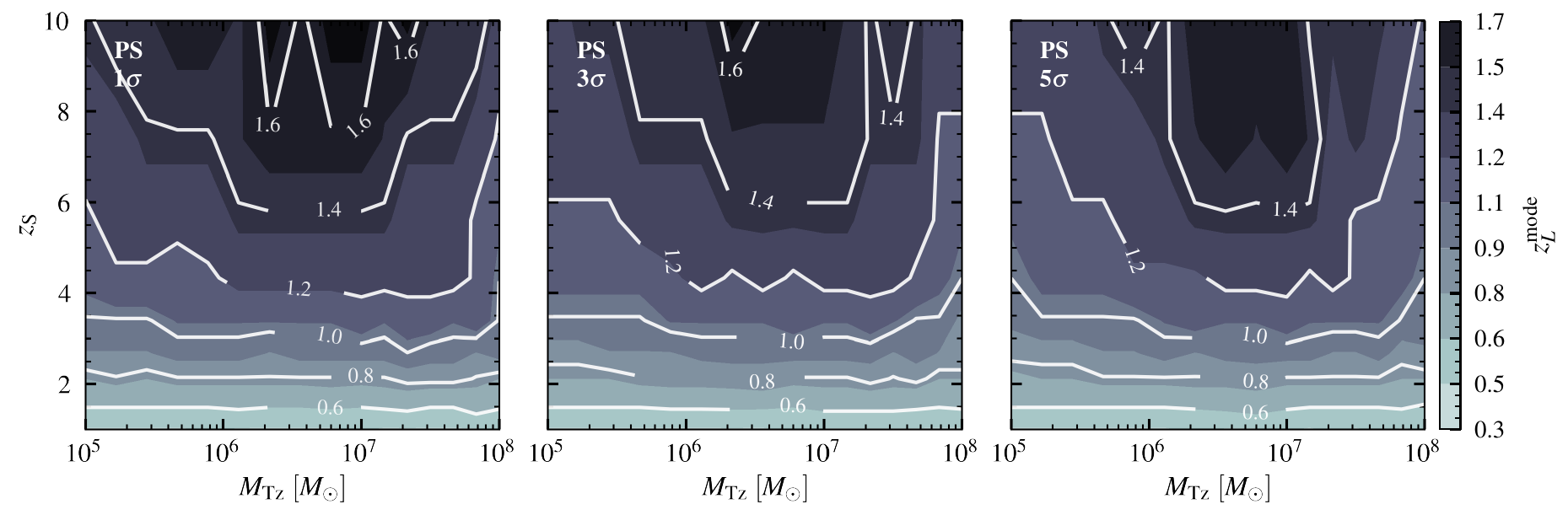}
    \caption{Mode of the differential optical depth per lens-redshift bin $z_{\rm L}^{\rm mode}$ assuming the PS halo mass function, as a function of $M_{\rm Tz} \in [10^5, 10^8]\ \Msun$ and $z_{\rm S} \in [1, 10]$).
      The panels refer to different confidence levels ($1\sigma$, $3\sigma$, $5\sigma$).
      The likeliest lens redshift is in the range $\left[0.3, 1.7\right]$, with a peak near $\sim 1.4$ for the MBHBs with the highest lensing probability (cf.~Fig.~\ref{fig:optical_depth_PS}).
    }%
    \label{fig:mode_Z_L_PS}%
\end{figure*}

The lowest possible values of $M_{\rm L}^{\rm prob}$  are particularly interesting, as they are usually hard to probe by other means.
These correspond to MBHBs with $M_{\rm Tz} \approx 10^6\, \textrm{--}\, 10^7\, M_\odot$, which have high SNRs. Binaries with $M_{\rm Tz} \gtrsim 5\times 10^7\, M_{\odot}$ produce longer-wavelength radiation and thus probe higher-mass lenses; lower-mass MBHBs, despite their potential sensitivity to lower lens masses, are penalized by their low SNRs.
Low-redshift MBHBs can probe lower lens masses (once again owing to their higher SNRs), but their optical depths are not large enough. Higher detection thresholds invariably result in higher $M_{\rm L}^{\rm prob}$ (compare the lower panels of Fig.~\ref{fig:2by2}).

In Fig.~\ref{fig:mode_ML_PS} of Appendix~\ref{App:lens_mass}, we show an alternative representation of these results. There we display the most probable lens mass that can be probed (i.e., the mode of the differential optical depth per lens mass bin) as a function of source parameters.

\subsubsection{Measured Velocity Function}

Figure~\ref{fig:M_L_inferred_MVF_allSigma} mirrors Fig.~\ref{fig:M_L_inferred_PS_allSigma}, but for the MVF lens population. 
There are two significant differences.

The first difference is that the $M_{\rm L}^{\rm prob}$ range is several orders of magnitude higher, because the WO optical depth has support at higher lens mass values in the MVF scenario
(see the lower-right panel of Fig.~\ref{fig:2by2}).

The second difference is that the 68\% confidence region of $M_{\rm L}^{\rm prob}$ in the MVF scenario is wider than in the PS case. This might seem counterintuitive at first, because the distributions in the lower panels of Fig.~\ref{fig:2by2} seem wider in the PS scenario. However, in Fig.~\ref{fig:2by2} we show the differential optical depths {\em per $\log_{10}$ lens-mass bin}, not the differential optical depth {\em per lens-mass bin}---hence the difference. 
The critical impact parameter peaks at low lens masses, which corresponds to the range where the comoving number density of objects is concentrated in the PS scenario. Therefore, the resulting differential optical depth distribution is narrower.
Conversely, in the MVF scenario, the comoving number density of low-mass lenses is lower, and the distribution is wider.

\subsubsection{Summary}

In summary, the most likely lens masses probed by LISA are within the range $M_{\rm L} \in (10^3, 10^8)\, M_{\odot}$, assuming the PS lens population. The sources where the WO probability is highest probe the lens mass range $M_{\rm L} \in (10^4, 10^6)\, M_{\odot}$ (cf. Fig.~\ref{fig:optical_depth_PS}). 

This is an important finding: LISA can explore (sub)halos with significantly lower masses, shedding light on the nature of DM in domain that is challenging to observe by other techniques.
The possibility to probe these lens masses within the duration of the mission, however, is strongly dependent on two poorly constrained quantities: the MBHB rate (see Table~\ref{tab:population_results}) and the underlying lens population (see Figs.~\ref{fig:optical_depth_PS} and~\ref{fig:optical_depth_MVF}).

\subsection{Halo Redshift Range Probed by LISA} \label{sec:results_lens_redshift}

We now turn to the range of lens redshifts that can be observed through WO effects. We apply the same methodology as in Sec.~\ref{sec:results_lens_mass}, but we are now interested in the most probable lens redshift for a specific MBHB, and so we focus on the mode of the differential optical depth per lens-redshift bin.

In Fig.~\ref{fig:mode_Z_L_PS}, we plot the most probable lens redshift for equal-mass MBHBs, as a function of $M_{\rm Tz}$ and $z_{\rm S}$. 
We assume the PS halo mass function and three different detection thresholds.
The most probable lens redshift is in the range $[0.3,\,1.7]$, with a peak around $\sim 1.2\, \textrm{--}\, 1.5$ for the MBHBs with the highest lensing probability (cf.~Fig.~\ref{fig:optical_depth_PS}).
Therefore, LISA can probe halos at fairly high redshifts, complementing SL and other observational techniques.
For brevity, we do not show results in the MVF scenario, where the WO probabilities are significantly lower.

\section{Conclusions} \label{sec:Conclusions}

The large-scale cosmic structure provides crucial information about the history of the Universe and its fundamental components, notably dark matter (DM).
Current techniques, such as strong lensing (SL), are limited to probing halos above $\sim 10^9\, \Msun$. 
Probing the substructures within DM halos remains challenging because of their faint luminosity and low masses, which lead to negligible detection probabilities.
However, low-mass DM substructures introduce new phenomenology. The wavelength of gravitational waves (GWs) can be comparable to the size of the lenses, leading to \textit{frequency-dependent} modulations in the waveform phase and amplitude due to wave-optics (WO) effects. 
At least in principle, this allows for the possibility to infer the lens properties (such as mass and profile) from single GW observations. In addition, WO effects can lead to lensing detectability with cross sections larger than those achievable with the observation of multiple ``images.''

The massive black hole binaries (MBHBs) detectable by LISA can have high SNRs, large redshifts, and low frequencies. As such, they are excellent candidates for observing WO effects. As long as MBHB rates are large enough, LISA can probe subhalo abundances, profiles, and masses, potentially testing DM formation scenarios.

We calculated the likelihood of observing WO effects with LISA through two distinct approaches: (i) a model-agnostic exploration of $\sim 1000$ MBHBs spanning the total mass, mass ratio, and redshift ranges relevant to LISA, and (ii) a study of three representative semi-analytical astrophysical models of MBHB populations.
This is (to our knowledge) the first study which 
(i) extensively investigates the influence of mass, mass ratio, redshift, and other source parameters on the observability of WO effects; (ii) calculates WO detection rates using astrophysical population models; (iii) considers both theory- and observation-based lens populations; (iv) examines different detection thresholds; and (v) employs the information matrix formalism using inspiral-merger-ringdown waveform models that include aligned spins and higher-order modes, factoring in potential parameter degeneracies and leveraging the full detector response of LISA. Two of our key predictions are the ranges of halo masses and redshifts that LISA can probe using WO effects.

\noindent
{\bf \em WO detection probability: model-agnostic approach.}
The model-agnostic calculation with a Press-Schechter (PS) lens population (Fig.~\ref{fig:optical_depth_PS}) yields WO detection probabilities between $0.5\%\, \textrm{--}\, 3\%$, $0.2\%\, \textrm{--}\, 2.0\%$, and $0.2\%\, \textrm{--}\, 1.0\%$ for $1\sigma$, $3\sigma$, and $5\sigma$ thresholds, respectively. The optimal sources for lensing observations are MBHBs with detector-frame total masses $M_{\rm Tz}\in [5\times 10^6\, \textrm{--}\, 5\times 10^7]\, \Msun$ at redshifts $z_{\rm S} \sim 5\, \textrm{--}\, 9$. 
In contrast, the measured velocity function (MVF) lens population yields probabilities that are an order of magnitude lower across the parameter space (Fig.~\ref{fig:optical_depth_MVF}). This observational lower bound indicates that while there is a non-negligible probability of observing WO effects with LISA, the rates depend heavily on the underlying MBHB source and lens populations.

The probabilities for unequal-mass binaries (shown in Figs.~\ref{fig:optical_depth_PS_q5} and~\ref{fig:optical_depth_PS_q10} of Appendix~\ref{App:Higher_Mass_Ratios})
are only $30\%$ to $50\%$ lower than the probabilities for equal-mass binaries, because the measurement improvement due to higher-order modes partly compensates for their lower SNR.

\noindent
{\bf \em Comparison with previous work.}
We can gain further insight by comparing our findings with prior work~\cite{Fairbairn:2022xln,Savastano:2023spl}.

A calculation using the mismatch-based Lindblom criterion and the \texttt{IMRPhenomD} waveform model~\cite{Savastano:2023spl} found values of $y_{\rm cr}$ slightly larger than those listed in Table~II of our previous work~\cite{Caliskan:2022hbu}, where we used an information matrix analysis based on the \texttt{IMRPhenomHM} model
and a $1\sigma$ detection threshold.
The difference would have been larger if we had used the \texttt{IMRPhenomD} model: this is because the \texttt{IMRPhenomHM} model, which includes higher-order multipoles, generally leads to higher SNR than the \texttt{IMRPhenomD} model, and higher-order multipoles reduce errors by removing parameter degeneracies~\cite{Caliskan:2022hbu}.
Similarly, the difference would have been larger if we had used stricter detection criteria (rather than the more optimistic $1\sigma$ threshold).
The detection probability estimated in Ref.~\cite{Savastano:2023spl} has a peak of $\sim 20\%$. This exceeds by one order of magnitude our most optimistic estimate, corresponding to a $1\sigma$ detection threshold and the PS lens population. The discrepancy approaches two orders of magnitude if we consider stricter detection criteria. In conclusion, our work suggests that the Lindblom criterion (by overlooking parameter degeneracies) overestimates WO detection probabilities by one or more orders of magnitude, depending on the chosen detection threshold.

On the contrary, our peak probability exceeds the estimates of Ref.~\cite{Fairbairn:2022xln} by over two orders of magnitude, because their study considers inspiral-only waveforms computed at lowest-order and a narrow halo mass range ($<10^8\, \Msun$). This comparison highlights the importance of including more realistic waveform models when estimating WO detection probabilities.

\noindent
{\bf \em WO detection probability: astrophysical population models.}
The most optimistic rates of events with observable WO effects estimated using astrophysical population models (see Table~\ref{tab:population_results} of Sec.~\ref{sec:source_pop_model}) correspond to the heavy seed (no delay) model combined with the PS lens population. In this case, we estimate $\sim 8$, $\sim 4$, and $\sim 3$ lensed events during a 4-year LISA mission for the $1\sigma$, $3\sigma$, and $5\sigma$ detection thresholds, respectively. The light seed scenario with the PS lens population is the only other case expected to yield at least one event at the $1\sigma$ threshold. The pessimistic, ``lower bound'' MVF scenario shows that it may be possible to observe no events with WO effects over the duration of the mission.

\noindent
{\bf \em Comparison with strong lensing.}
The model-agnostic calculation yields probabilities of observing WO effects that generally exceed the SL probabilities.
While WO and SL probabilities are similar at high redshifts, WO probabilities are larger (even if we assume strict detection thresholds) at intermediate and lower redshifts.

For the astrophysical population models, we find that WO lensing probabilities for the PS lens population are significant. 
The probabilities for both heavy seed scenarios are in the range $0.5\%\, \textrm{--}\, 1.5\%$ for all detection thresholds. 
This rate is higher than the SL rates of quasars~\cite{Treu:2005aw, Saha:2006cx, 2012AJ....143..119I} and supernovae~\cite{Kelly:2014mwa, 2018NatAs...2..324R, Goobar:2016uuf}, and higher than the anticipated lensing rate for stellar-mass black hole binaries~\cite{Xu:2021bfn, LIGOScientific:2023bwz}.

Our rate estimates in the PS case are larger than the WO and SL rates found in Ref.~\cite{Sereno:2010dr}, probably because those early estimates overlooked WO effects. 
On the contrary, our estimated WO rates are significantly lower than SL rates in the MVF scenario.

\noindent
{\bf \em Range of halo masses and redshifts accessible to LISA.}
In Fig.~\ref{fig:M_L_inferred_PS_allSigma}, we compute the probable range of lens masses $M_{\rm L}^{\rm prob}$ as a function of the source parameters. We find that most likely lens mass is in the range $(10^3, 10^8)\, \Msun$, and that the probed lens mass range is approximately $M_{\rm L} \in (10^4, 10^6)\, \Msun$ for sources which have the highest WO probability (see Fig.~\ref{fig:optical_depth_PS}).
Therefore, LISA can explore (sub)halos with masses significantly lower than those currently achievable with other techniques.
If there is a dearth of low-mass lenses (as in the MVF scenario of Fig.~\ref{fig:M_L_inferred_MVF_allSigma}), this may not be possible.

The most likely lens redshifts accessible by LISA (Fig.~\ref{fig:mode_Z_L_PS}) are in the range $(0.3, 1.7)$, peaking around $\sim 1.2\, \textrm{--}\, 1.5$ for MBHBs with the highest probability: LISA can probe (sub)halos at comparatively high redshifts, complementing other observational techniques.

\noindent
{\bf \em Future research directions.}
We modeled lenses by creating look-up tables for singular isothermal spheres.
It will be interesting to extend our rapid parameter estimation techniques to more complex lenses and compute the corresponding rate estimates.
The ability of various detectors to distinguish between different lens models in statistical analyses should be estimated by taking into account detection probabilities and by developing more advanced parameter estimation techniques.

Our results could be extended to new MBHB population models as our understanding of MBHBs improves, e.g., through observations by the James Webb Space Telescope~\cite{Volonteri:2022yhe,Bogdan:2023ilu} and pulsar timing arrays (PTAs)~\cite{Steinle:2023vxs,NANOGrav:2023hfp,Ellis:2023dgf}.

Our study considers various lens populations, but the lensing probabilities are evaluated for isolated halos.
While it is unlikely to have multiple lenses with observable WO effects affecting a particular MBHB, the development of parameter estimation techniques should not be confined to individual lenses. 
 Future work should incorporate the effects of subhalo clustering within parent halos. It is also important to understand the potential impact and possible detections of line-of-sight halos~\cite{Sengul:2021lxe}.

The observation of lensed GWs with WO effects is not limited to LISA. 
Our techniques can be extended to assess detection rates for ground-based detectors, which are sensitive to the lower-mass end of the halo mass function and to a different population of compact objects.
Another compelling avenue for future exploration involves detecting WO effects in anticipated, resolvable PTA sources. 
These low-frequency ($\sim$nHz) GW sources would be sensitive to WO effects from halos with virial masses up to $M_{200} \sim (10^{12}\, \textrm{--}\, 10^{14})\, \Msun$, offering unprecedented insights into galaxy-scale (or larger) structures.

\begin{figure*}[t]
    \includegraphics[width = \linewidth]{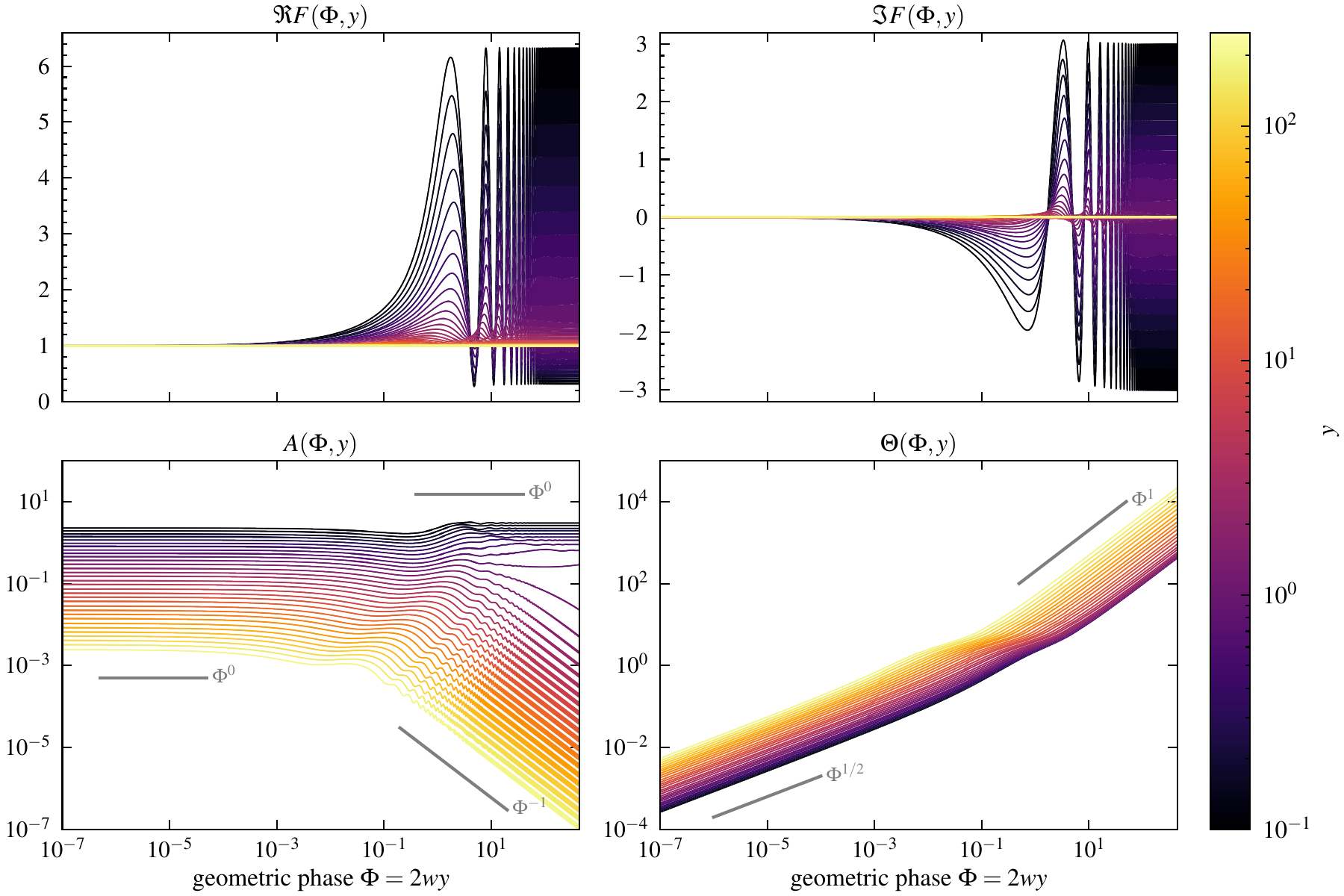}
    \caption{Lensing diffraction integral before (top panels) and after (bottom panels) the regularizing transformation defined in Eqs.~\eqref{eqn:O-function}, \eqref{eqn:A-function}, and \eqref{eqn:Theta-function}. Here, the geometric phase $\Phi \equiv 2wy$ is used instead of the dimensionless frequency $w$ for convenience. 
    Different colors represent different values of the impact parameter $y$. 
    Gray line segments are used as references for the relevant power-law indices.}
    \label{fig:lut}
\end{figure*}

\acknowledgements
We thank Charles R.~Keeton, Sylvain Marsat, and Anowar J.~Shajib for valuable feedback and comments.
E.B.~and M.\c{C}.~are supported by NSF Grants No.~AST-2006538, PHY-2207502, PHY-090003 and PHY-20043, and NASA Grants No. 20-LPS20-0011 and 21-ATP21-0010. 
E.B.~acknowledges support from the ITA-USA Science and Technology Cooperation programme (CUP: D13C23000290001), supported by the Ministry of Foreign Affairs of Italy (MAECI).
M.K.~was supported by NSF Grant No. 2112699 and the Simons Foundation.
M.\c{C}.~is also supported by Johns Hopkins University through the Rowland Research Fellowship.
J.M.E.~is supported by the European Union’s Horizon 2020 research and innovation program under the Marie Sklodowska-Curie grant agreement No.~847523 INTERACTIONS, and by VILLUM FONDEN (grant no.~53101 and 37766).
This work was carried out at the Advanced Research Computing at Hopkins (ARCH) core facility (\url{rockfish.jhu.edu}), which is supported by the NSF Grant No.~OAC-1920103. 
The authors acknowledge the Texas Advanced Computing Center (TACC) at The University of Texas at Austin for providing {HPC, visualization, database, or grid} resources that have contributed to the research results reported within this paper~\cite{10.1145/3311790.3396656}. URL: \url{http://www.tacc.utexas.edu}

\appendix

\section{Evaluation of the Lensing Diffraction Integral Using Look-up Tables}\label{App:LUT}

The numerous information-matrix analyses required to explore the detectable parameter space demand efficient evaluation of the diffraction integral $F(w, y)$. 
For the SIS density profile, the diffraction integral is not only tricky, but also expensive to evaluate directly~\cite{Guo:2020eqw}.
A naive interpolation of the numerical results suffers from (i)~needing a large number of data points to resolve the rapid oscillations in $w$, and (ii)~ambiguities in extrapolating outside of the $w$ range of direct evaluation. 
Here, we devise a simple analytical regularizing transformation of $F(w, y)$ so that it has only mild oscillations in $w$, as well as a simple scaling behavior for $w \gg 1$ and $w \ll 1$.

Consider the mean-zero oscillatory part of $F(w, y)$,
\begin{equation}\label{eqn:O-function}
    O(w, y) \equiv F(w, y) - |\mu_+(y)|^{1/2},
\end{equation}
where $\mu_+(y) \equiv 1 + 1/y$ is the geometric-optics magnification of the Type-I image~\cite{Schneider:1992}.

We note that the geometric-optics approximation of $F(w,  y)$ has oscillations of period $\pi / y$ in $w$ for $y \leq 1$, so we introduce the geometric phase $\Phi \equiv 2wy$ and use it in place of $w$ (for a given $y$) to scale the oscillations to having period $2\pi$ in $\Phi$. We define the amplitude,
\begin{equation}\label{eqn:A-function}
    A(\Phi, y) \equiv |O[w = \Phi / (2y), y]|,
\end{equation}
and the phase,
\begin{equation}\label{eqn:Theta-function}
    \Theta(\Phi, y) \equiv \pi + \arg\, O[w = \Phi / (2y), y],
\end{equation}
of $O(w, y)$ so that the oscillatory part $O(w, y)$ can be reconstructed as
\begin{equation}\label{eqn:O-reconstruction}
    O(w, y) = A(\Phi = 2wy, y) \exp[i \Theta(\Phi = 2wy, y) - i\pi],
\end{equation}
and $F(w, y)$ as
\begin{equation}\label{eqn:F-reconstruction}
    F(w, y) = O(w, y) + |\mu_+(y)|^{1/2}.
\end{equation}
Therefore, the real-valued functions $A(\Phi, y)$ and $\Theta(\Phi, y)$ encode the same information as $F(w, y)$. Here, $\arg z \equiv -i \ln (z / |z|)$ is the argument of the complex number $z$, where the correct branch of the $\ln$ function is chosen to give a continuous argument when $z$ is varying continuously.
In the definition of $\Theta(\Phi, y)$, we offset the argument by $\pi$ so that $\Theta(\Phi, y) \to 0_+$ when $\Phi \to 0$. Under these definitions, $A(\Phi, y)$ and $\Theta(\Phi, y)$ are both positive.

\begin{figure*}[t!]
    \centering
    \includegraphics[width=.99\textwidth]{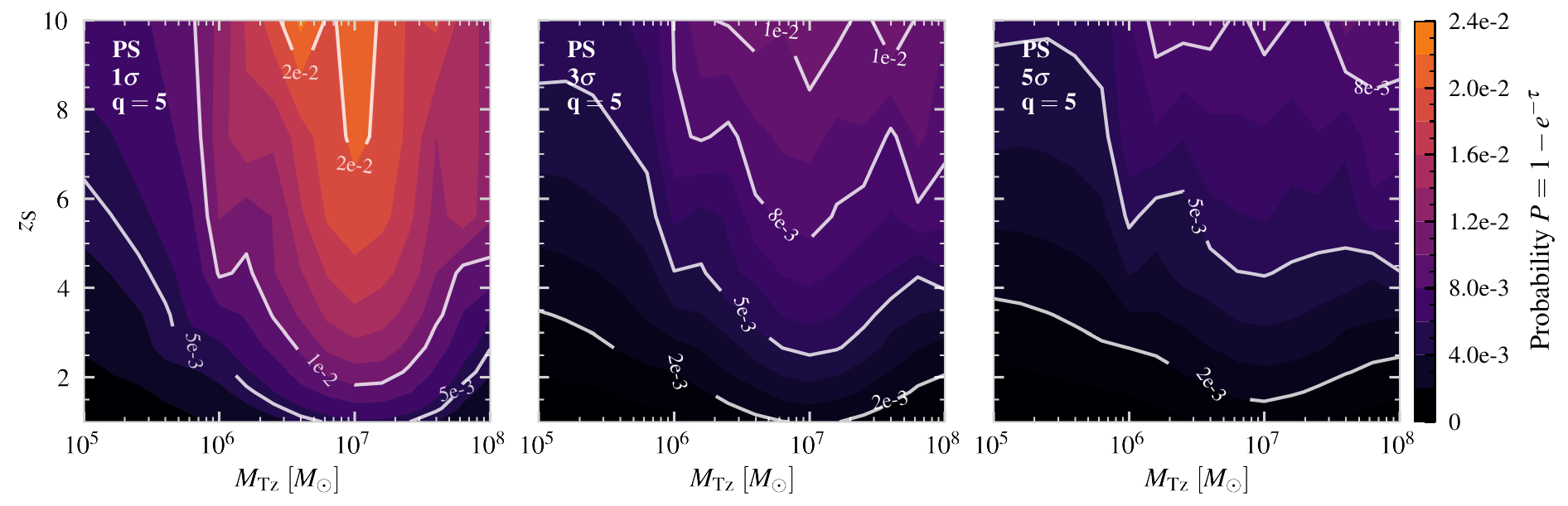}
    \caption{Same as Fig.~\ref{fig:optical_depth_PS}, but for mass ratio $q=5$.
    }%
    \label{fig:optical_depth_PS_q5}%
\end{figure*}

\begin{figure*}[t!]
    \centering
    \includegraphics[width=.99\textwidth]{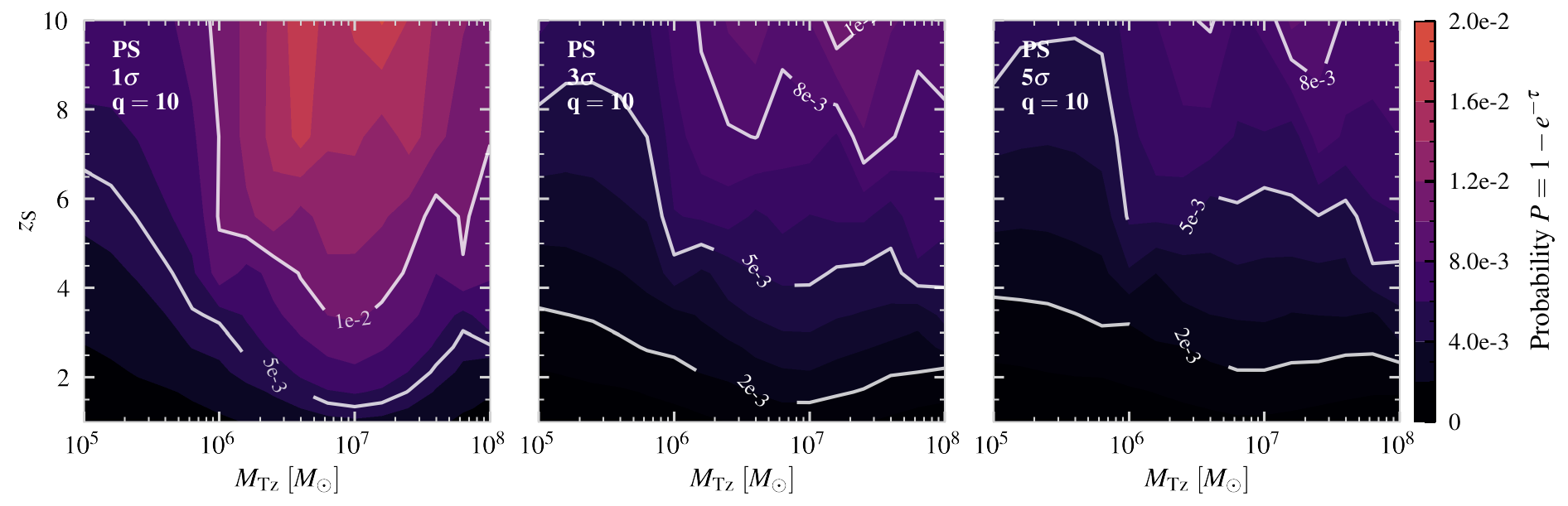}
    \caption{Same as Fig.~\ref{fig:optical_depth_PS}, but for mass ratio $q=10$.
    }%
    \label{fig:optical_depth_PS_q10}%
\end{figure*}

In Fig.~\ref{fig:lut} we show the diffraction integral $F(\Phi, y) \equiv F[w = \Phi / (2y), y]$ obtained in the same manner as detailed in the Appendix~A of Ref.~\cite{Caliskan:2022hbu} (top panels) and the functions $A(\Phi, y)$ and $\Theta(\Phi, y)$ defined in Eqs.~\eqref{eqn:A-function} and \eqref{eqn:Theta-function}, respectively (bottom panels).

The oscillatory complex-valued function $F(\Phi, y)$ is reduced to two real-valued functions $A(\Phi, y)$ and $\Theta(\Phi, y)$, with only mild features and power-law scaling behaviors outside the range of direct evaluation. Note that the power-law index for $A(\Phi, y)$ for large $\Phi$ is different for $y \leq 1$ ($A \propto \Phi^0$) and for $y > 1$ ($A \propto \Phi^{-1}$). We interpolate the functions $A(\Phi, y)$ and $\Theta(\Phi, y)$ using bivariate cubic splines in $\Phi$ and $y$ (in log-log space) within the range of direct evaluation, and extrapolate in $\Phi$ using the appropriate power-law index.  No extrapolation in $y$ is needed for our application.  We reconstruct $F(w, y)$ using Eqs.~\eqref{eqn:O-reconstruction} and \eqref{eqn:F-reconstruction}.  We also reconstruct the partial derivatives $\partial_w F(w, y)$ and $\partial_y F(w, y)$ needed for the information-matrix analysis via the appropriate chain rules from the spline derivatives $\partial_\Phi A(\Phi, y)$, $\partial_y A(\Phi, y)$, $\partial_\Phi \Theta(\Phi, y)$, and $\partial_y \Theta(\Phi, y)$.

\begin{figure*}[t]
    \centering
    \includegraphics[width=.99\textwidth]{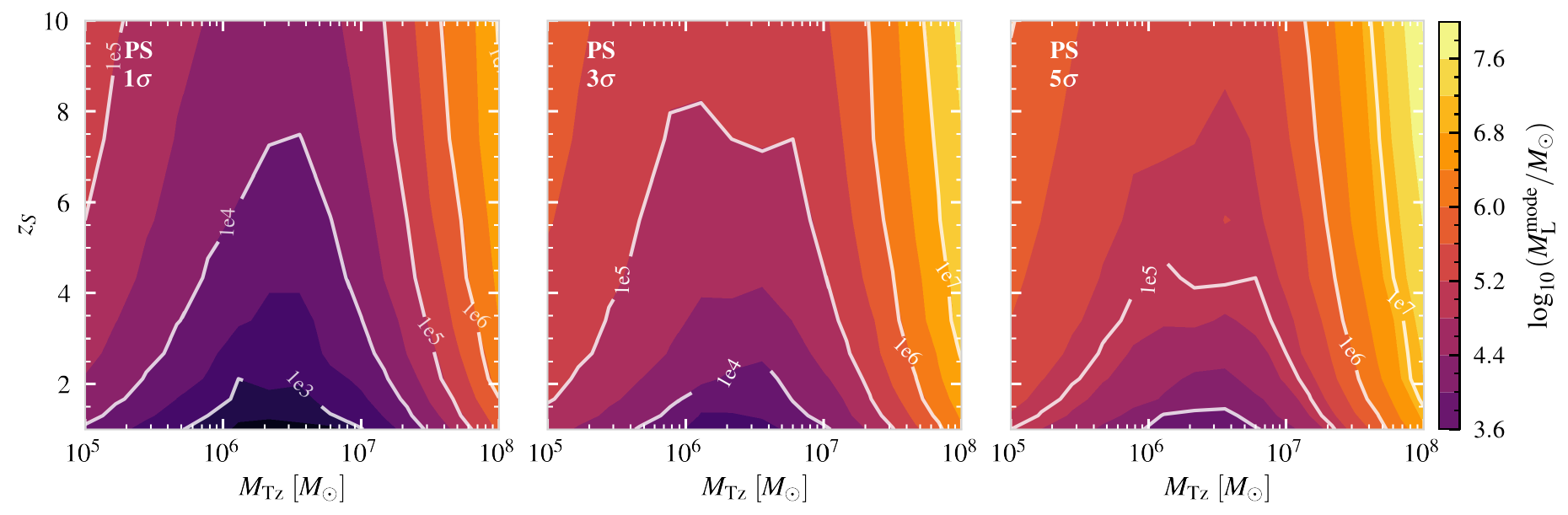}
    \caption{The mode of differential optical depth per lens-mass bin, $M_{\rm L}^{\rm mode}$, according to the PS halo mass function.
    The contour plots show the most likely lens mass for each source under various confidence levels ($1\sigma$, $3\sigma$, $5\sigma$) and source parameters, with $M_{\rm Tz} \in [10^5, 10^8]\ \Msun$ and $z_{\rm S} \in [1, 10]$.
    The most likely lens mass is in the range $\left[10^3, 10^8\right]\, \Msun$, and it is $\sim 10^5\, \Msun$ for the MBHBs with the highest lensing probability (cf.~Fig.~\ref{fig:optical_depth_PS}).
    }%
    \label{fig:mode_ML_PS}%
\end{figure*}

\begin{figure*}[t]
    \centering
    \includegraphics[width=.99\textwidth]{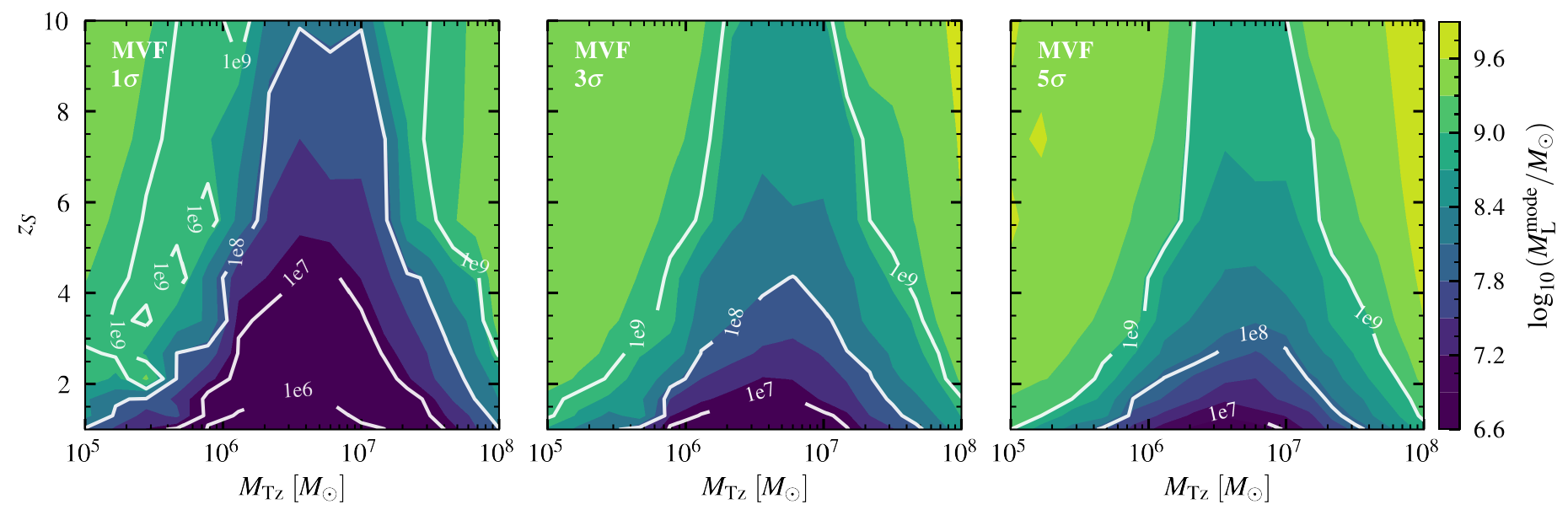}
    \caption{Same as Fig.~\ref{fig:mode_ML_PS}, but for the MVF lens population. 
    The most likely lens mass ranges between $\left[10^6, 10^{10}\right]\, \Msun$, and it is $\sim 10^{10}\, \Msun$ for the MBHBs with the highest lensing probability (cf.~Fig.~\ref{fig:optical_depth_MVF}).
    }%
    \label{fig:mode_M_L_MVF}%
\end{figure*}

\begin{figure}[h!]
\includegraphics[width=\columnwidth]{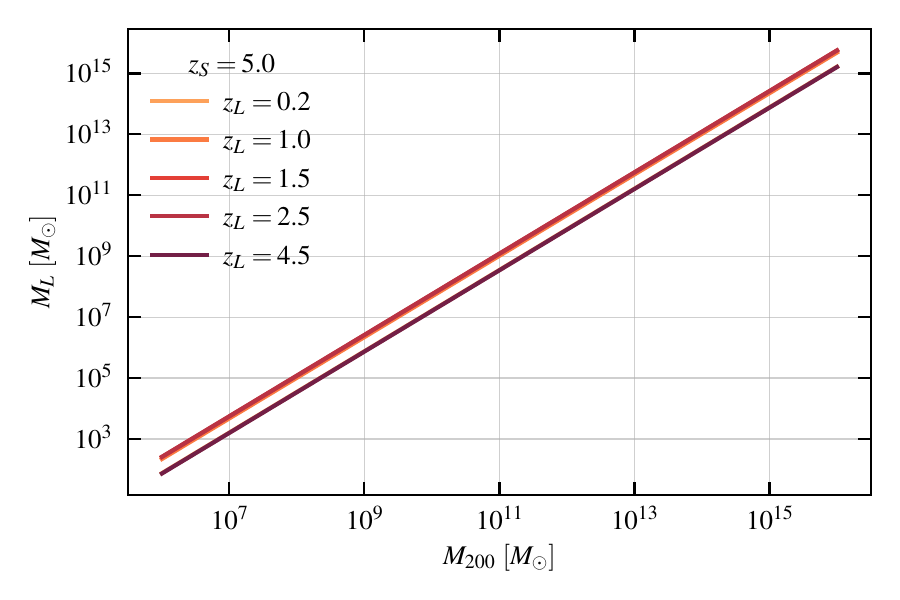} 
\caption{The relationship between the lens mass $M_{\rm L}$ and the halo virial mass $M_{200}$ for various lens redshifts, assuming an SIS mass-density profile and a source at redshift $z_{\rm S} = 5$.}
\label{fig:Mvir_Mlens}
\end{figure}

\section{Higher Mass Ratios}\label{App:Higher_Mass_Ratios}

In this section, we extend the discussion of Sec.~\ref{sec:results_model_agnostic} to MBHBs with unequal masses.

Figures~\ref{fig:optical_depth_PS_q5} and~\ref{fig:optical_depth_PS_q10} are similar to Fig.~\ref{fig:optical_depth_PS}, but for MBHBs with mass ratio $q=5$ and $q=10$, respectively.
We restrict our investigation to $q \leq 10$ because, according to astrophysical models, most MBHBs should have comparable masses (see Fig.~\ref{fig:source_pop}).
The lensing probabilities are marginally reduced (by 30\% to 50\%) compared to equal mass binaries, suggesting that lensing rates do not drastically decrease with $q$, so the majority of MBHBs detectable by LISA have lensing probabilities within the range of $0.02\%\, \textrm{--}\, 2\%$ at the $3\sigma$ detection threshold.

We do not show results for unequal-mass MBHBs in the MVF scenario because the probabilities for equal-mass binaries are already insignificant in this case.

\section{Mode of Probed Halo Masses} \label{App:lens_mass}

In Sec.~\ref{sec:results_lens_mass} we compute the range of lens masses that LISA can probe by showing the median (and the 68\% confidence interval) of the differential optical depth per lens-mass bin as a function of $M_{\rm Tz}$ for selected values of $z_{\rm S}$. Here we present the most likely lens mass (i.e. the mode of the differential optical depth, which provides the lens mass associated with the peak probability in the distribution) as a function of $M_{\rm Tz}$ and $z_{\rm S}$. These results are complementary to the contour plots in Figs.~\ref{fig:optical_depth_PS} and~\ref{fig:optical_depth_MVF}.

In Fig.~\ref{fig:mode_ML_PS} we plot the most probable lens mass as a function of MBHB parameters for the PS halo mass function and different detection thresholds.
The most likely lens mass is in the range $(10^{3}, 10^{8})\, M_\odot$, and it is around $(10^4, 10^6)\, M_{\odot}$ when the lensing probability attains its maximum (refer to Figs.~\ref{fig:optical_depth_PS},~\ref{fig:optical_depth_PS_q5}, and~\ref{fig:optical_depth_PS_q10}), contingent on the detection threshold.
Figure~\ref{fig:mode_M_L_MVF} is analogous to Fig~\ref{fig:mode_ML_PS}, but for the MVF lens population. 
The most likely lens mass is now in the range $(10^{6}, 10^{10})\, M_\odot$, and it is around $10^9\, M_{\odot}$ when the lensing probability peaks (refer to Fig.~\ref{fig:optical_depth_MVF}). 
This shift relative to the PS case is anticipated: the MVF population has a smaller number density of low-mass lenses, leading to WO optical depths that are more widely distributed and peak at relatively higher masses. 
The correlation between lens mass and halo virial mass for a few source/lens redshift configurations is given, for reference, in the Appendix~\ref{App:lens_mass_vs_halo_mass}.

\newpage
\section{Relationship Between Lens Mass and Halo (Virial) Mass} \label{App:lens_mass_vs_halo_mass}

In Figure~\ref{fig:Mvir_Mlens} we present the relationship between the halo virial mass $M_{200}$ and the lens mass $M_{\rm L}$ for selected lens redshifts and for an SIS mass-density profile (see Sec.~\ref{sec:mass_function_to_profile}). 
The redshift at the source is set at $z_{\rm S} = 5$.

\clearpage
\bibliography{LISA_Lensing_Population}

\end{document}